\documentclass[a4paper,11pt]{article}
\pdfoutput=0
\usepackage{jheppub}

\usepackage[utf8]{inputenc}
\usepackage{graphicx}
\usepackage{float}
\usepackage{amssymb}
\usepackage{amsmath}  
\usepackage{dsfont}
\usepackage{color}
\usepackage{array}
\usepackage[locale=US]{siunitx}	
\usepackage{bm}
\usepackage{mathrsfs}
\usepackage{pifont}

\newcommand{\X}{\ding{53}}

\title{Non-Abelian string breaking phenomena with Matrix Product States}
\author{Stefan K\"uhn,}
\author{Erez Zohar,}
\author{J. Ignacio Cirac,}
\author{and Mari Carmen  Ba\~nuls}

\affiliation{Max-Planck-Institut f\"{u}r Quantenoptik, Hans-Kopfermann-Str. 1, D-85748 Garching, Germany}

\emailAdd{stefan.kuehn@mpq.mpg.de}
\emailAdd{erez.zohar@mpq.mpg.de}
\emailAdd{ignacio.cirac@mpq.mpg.de}
\emailAdd{banulsm@mpq.mpg.de}

\abstract{Using matrix product states, we explore numerically the phenomenology of string breaking in a non-Abelian lattice gauge theory, namely 1+1 dimensional SU(2). The technique allows us to study the static potential between external heavy charges, as traditionally explored by Monte Carlo simulations, but also to simulate the real-time dynamics of both static and dynamical fermions, as the latter are fully included in the formalism. We propose a number of observables that are sensitive to the presence or breaking of the flux string, and use them to detect and characterize the phenomenon in each of these setups.}

\keywords{Field Theories in Lower Dimensions, Lattice Gauge Field Theories}

\arxivnumber{1505.04441}

\begin{document}
\maketitle
\section{Introduction}
The phenomenon of string breaking, as a consequence of confinement, is one of the most fundamental aspects of the gauge theories that 
lie at the basis of our understanding of high energy physics. While a comprehensive understanding in many cases, such as QCD, is still lacking, a lot of insight has been gained thanks to lattice gauge theory (LGT). In LGT, originally pioneered by Wilson \cite{Wilson1974}, the theory is formulated on a discretized space-time lattice in such a way that the gauge symmetry is preserved and the continuum model is recovered in the limit of vanishing lattice spacing. This allows powerful Monte Carlo simulations which besides mass spectra \cite{Durr2008} and phase  diagrams \cite{Fukushima2011} have  been also successfully used to study static properties of confinement and of string breaking \cite{Bali1992,Philipsen1998,Knechtli1998,Chernodub2002,Bali2005,Pepe2009}. Despite the great success of this method, it suffers from the sign problem  \cite{Troyer2005}, which limits the kind of feasible simulations, and in particular real-time dynamics (although recently introduced techniques have allowed some real-time calculations in certain regimes \cite{Hebenstreit2013a,Hebenstreit2013}). It is then of undeniable interest to explore other techniques that can overcome the limitations of standard LGT techniques.

A particularly promising approach is the application of tensor networks (TN) to the Hamiltonian formulation of LGT \cite{Byrnes2002,Banuls2013,Banuls2013a,Buyens2013,Rico2013,Tagliacozzo2013,Tagliacozzo2014,Silvi2014,Buyens2014,Saito2014,Banuls2015,Tagliacozzo2011}. These methods are free from the sign problem and allow the computation of real-time evolution \cite{Vidal2003,Verstraete2008,Daley2004}, therefore opening up a new perspective of addressing dynamics of problems that are currently intractable with Monte Carlo simulations. The first numerical results for static properties \cite{Banuls2013,Banuls2013a,Buyens2013,Buyens2014}, finite temperature calculations \cite{Saito2014,Banuls2015} and out-of-equilibrium dynamics \cite{Buyens2013,Buyens2014} indeed show the power of the method.

Another direction explored in recent years is quantum simulation of gauge theories \cite{Cirac2010,Zohar2011,Zohar2012,Banerjee2012,Banerjee2013,Hauke2013,Rico2013,Stannigel2013,Tagliacozzo2013a,Wiese2013,Zohar2013,Zohar2013a,Zohar2013b,Zohar2013c,Kosior2014,Kuehn2014,Marcos2014,Wiese2014,Mezzacapo2015,Notarnicola2015,Zohar2015a}. The proposals for quantum simulators typically work as well with the Hamiltonian formulation of the LGT and map the gauge theory to the physical degrees of freedom of another quantum system which can be controlled experimentally thus being free from purely numerical limitations. Although a fully fledged quantum simulator is not available yet, this route seems promising for the future as the first proposals might come into reach with nowadays high level of experimental control.

In this work, we consider a non-Abelian lattice gauge theory with dynamical charges in one spatial dimension. The Hamiltonian we simulate realizes an exact SU(2) gauge symmetry using a finite-dimensional representation for the bosonic degrees of freedom \cite{Zohar2015,Zohar2015a}. This model is hence a suitable candidate for the design of an atomic quantum simulator of SU(2) LGT. The finite dimension of the link Hilbert spaces allows the simulation of the model via direct application of MPS techniques, which we employ here in order to numerically explore the statical and dynamical aspects of the string breaking phenomenon. 

In order to detect and characterize the breaking of the string, we propose three different observables. They can be monitored during the MPS simulation, and should in principle also be accessible in an experiment, so that our results will be measurable in a potential future quantum simulator. 

More specifically we study three different scenarios. First, we consider the static aspects of string breaking by determining the ground state of the system, which includes fully dynamical fermions, in the presence of two additional external static charges. This calculation relates closely to the ones traditionally accessible by lattice Monte Carlo methods. We show that we can reliably determine if the string is present, and therefore identify the regions where we expect string breaking to occur. These calculations additionally demonstrate the suitability of the proposed observables for the detection of string breaking in dynamical scenarios. In the second place, we investigate the dynamics of string breaking by introducing the external static charges on top of the interacting vacuum and evolving the state in real time. When the string breaks, we can explicitly observe the screening of the charges via the creation of dynamical particles. Finally, we analyze how the picture changes when the charges added to the vacuum are themselves dynamical,  a scenario which is closer to more realistic out-of-equilibrium situations. Also in this case we can recognize the string breaking if the fermion mass is small enough.

The rest of the paper is organized as follows. In section \ref{sec:methods} we review the model and describe the numerical methods we are applying, with a special focus on the MPS techniques we are using. In section \ref{sec:static_qq} we present our numerical results for the static calculations. Section \ref{sec:dynamic_qq} contains the results for the real-time dynamics of string breaking with two static external charges. In section \ref{sec:dynamic_string} we present the results for the real-time dynamics of string breaking when a pair of dynamical charges is added to the interacting vacuum. To conclude, we summarize our findings in section \ref{sec:conclusions}.

\section{\label{sec:methods}Model and Methods}
\subsection{Truncated theory}

The model we consider is the Hamiltonian formulation of a 1+1 dimensional SU(2) LGT with dynamical fermions, in which the gauge symmetry is exactly realized with finite-dimensional link variables \cite{Zohar2015a}. The model can be understood as a truncation of the complete SU(2) gauge theory, formulated as the Kogut-Susskind Hamiltonian with staggered fermions \cite{Kogut1975}
\begin{align}
\begin{aligned}
H = &\varepsilon\sum_n \left(\psi_n^\dagger U_{n}\psi_{n+1}+\mathrm{h.c.}\right) \\
    &+ m\sum_n (-1)^n\psi_n^\dagger\psi_n+\frac{g^2}{2}\sum_n \mathbf{J}_{n}^2,
\end{aligned}
    \label{KogutSusskindHamiltonian}
\end{align}
where $\psi_n=\begin{pmatrix}\psi_{r,n}\\ \psi_{g,n} \end{pmatrix}$ is a two component spinor taking into account the two ``colors'' of fermions (``red'' and ``green'') on site $n$. $U_n$ is a matrix on color space in the fundamental representation. Its entries, $U_n^{ij}$, act on the link between the fermionic sites $n$ and $n+1$ and change the corresponding electric flux. The operator $\mathbf{J}^2_n$ is a group scalar, as it does not carry any color index, and gives the flux energy on link $n$. Physical states, $|\phi\rangle$, have to be eigenstates of Gauss Law, i.e. $G^\alpha_n|\phi\rangle=q^\alpha_n|\phi\rangle$ for all sites, $n$, where 
\begin{align*}
 G^\alpha_n &= L^\alpha_{n}-R^\alpha_{n-1}-Q^\alpha_{n},\quad\alpha\in\{x,y,z\}.
\end{align*}
$Q^\alpha_{n}=\frac{1}{2}\psi_n^\dagger\sigma^\alpha\psi_n$ is the non-Abelian dynamical charge at site $n$, where $\sigma^\alpha$ are the usual Pauli matrices. A nonzero value for $q^\alpha_n$ indicates a non-vanishing static external charge.\footnote{We are not considering these external charges explicitly in the Hamiltonian. As they are static, they would only contribute a constant mass term to eq. (\ref{KogutSusskindHamiltonian}) leading to a constant offset in energy in all our simulations, thus being negligible.} $L^\alpha_n$, $R^\alpha_n$ are the generators of left and right gauge transformations acting on link $n$, which correspond to the left and right electric field on the link and fulfill the commutation relations $[L^\alpha_n,L^\beta_m]=-i\delta_{nm}\sum_\gamma\varepsilon^{\alpha \beta \gamma} L^\gamma_n$, $[R^\alpha_n,R^\beta_m]=i\delta_{nm}\sum_\gamma\varepsilon^{\alpha \beta \gamma} R^\gamma_n$ and $[L^\alpha_n,R^\beta_m]=0$. They are related to the operator for the electric flux energy as $\mathbf{J}^2_n=\sum_\alpha L^\alpha_nL^\alpha_n = \sum_\alpha R^\alpha_nR^\alpha_n$. The Gauss Law components do not commute among themselves, as $[G^\alpha_n,G^\beta_m]=-i\delta_{nm}\sum_\gamma\varepsilon^{\alpha \beta \gamma} G^\gamma_n$, and thus cannot be diagonalized simultaneously. However, they all commute with the Hamiltonian, $[H,G^\alpha_n]=0$, so that the Hilbert space is a direct sum of sectors characterized by their configuration of external charges, $\{q^\alpha_n\}$ \cite{Zohar2015a}.   

In the strong coupling limit, corresponding to $\varepsilon=0$, the ground state is given by links carrying no flux, odd sites occupied with two fermions, and empty even sites. This state can be written in a product basis as
\begin{align*}
 |\phi_\mathrm{SC}\rangle = |\mathbf{2}\rangle|0\rangle|\mathbf{0}\rangle|0\rangle|\mathbf{2}\rangle|0\rangle\dots.
\end{align*}
In the previous expression bold numbers represent the occupation number of a fermionic site, while $|0\rangle$ represents the state of a link with no flux. The fermionic part of this state corresponds to the Dirac sea, and it is easy to check that it fulfills Gauss Law with $G^\alpha_n|\phi_\mathrm{SC}\rangle=0$ $\forall n,\alpha$. Applying on this state the gauge invariant string operator
\begin{align}
 S_{nl} = \psi^\dagger_n U_n \dots U_{n+l-1} \psi_{n+l}
 \label{string_op}
\end{align}
or its adjoint, $S^\dagger_{nl}$, for odd $l$,\footnote{As we are working with a staggered formulation the odd (even) sites correspond to antiparticles (particles), consequently $l$ has to be odd to create a string between an antiparticle and a particle.}  generates a particle-antiparticle pair at sites $n$ and $n+l$, connected by a flux tube of length $l$. Such configurations will have an excess of energy of $2m$ due to the particle-antiparticle pair, plus a flux energy proportional to the string length, $l$. Consequently, from a certain length on, it will be energetically favorable to reduce the flux energy by creating extra particle-antiparticle pairs, leading to configurations with a broken string. 

In the complete SU(2) model, the flux on a link is not bounded, and the dimension  of the Hilbert space for each link is infinite. It is nevertheless possible to consider a theory truncated in a gauge invariant manner \cite{Zohar2015}, where the maximum flux a link can carry is limited and the Hilbert spaces of the links are finite dimensional. Following the method in Ref. \cite{Zohar2015}, each matrix element $U_n^{ij}$ can be decomposed as a sum over all irreducible representations, which may be separated to summands that are gauge invariant themselves. Here we consider the model corresponding to the simplest non-trivial truncation of the full theory, meaning that only the trivial and the fundamental representation are kept, resulting in dimension $5$ for the links.

\subsection{Numerical Methods}

We use the MPS ansatz to describe the system \cite{Perez-Garcia2007}. A MPS with open boundary conditions for $N$ sites is given by
\begin{align*}
 |\Psi\rangle=\sum_{i_1,i_2,\dots, i_N}A^{i_1}_1A^{i_2}_2\dots A^{i_N}_N|i_1\rangle|i_2\rangle\dots|i_N\rangle,
\end{align*}
where the $A^{i_k}_k$ are complex matrices in $\mathds{C}^{D\times D}$ for $1<k<N$ and $A^{i_1}$ ($A^{i_N}$) is a row (column) vector. The states $|i_k\rangle_{i_k=1}^{d_k}$ form a basis of the $d_k$-dimensional local Hilbert space on site $k$. The number $D$, called bond dimension of the MPS, determines the number of variational parameters and limits the amount of entanglement that can be present in the MPS.

The string breaking phenomenon can be studied statically or dynamically. Different well known numerical algorithms exist to find MPS descriptions of stationary or time-dependent states \cite{Verstraete2008,Schollwoeck2011}. In this work, we employ simulations of time evolution \cite{Vidal2003,Verstraete2004a,Verstraete2008} where the evolution operator is approximated via a first order Taylor expansion. This allows us to study both static and dynamic scenarios by respectively using imaginary or real time. Additionally, the Taylor expansion preserves the symmetries of the Hamiltonian, such as gauge invariance. Then it is possible to explore a specific sector of the external charge distribution without explicitly implementing symmetries in the tensors.

The numerical simulations have three main sources of error, each of them controllable by a suitable choice of parameters. The first one is due to the approximation of the evolution operator via a Taylor series. This error can be controlled by choosing the time step suitably small. Another source of error is due to the limited bond dimension. By obtaining results with different bond dimensions, the size of this truncation error can also be estimated and controlled. As we are working with finite systems, a third source of error arises from finite size effects which can be avoided by using sufficiently large systems.

Although TN and in particular MPS can be formulated in terms of fermionic degrees of freedom, we choose to translate the fermionic degrees of freedom in Hamiltonian (\ref{KogutSusskindHamiltonian}) to spins via a Jordan-Wigner transformation for our numerical simulations (details about the procedure and the relevant operators in spin formulation can be found in the appendix \ref{app:spin_formulation}) and in the following we work with the spin formulation.

\subsection{Detection of string breaking}

Detecting the presence or absence of the flux string during the evolution requires observables that are suitable for the dynamical case. The  Wilson loop and more general correlation functions, widely used in lattice calculations to determine the static potential, are typically evaluated in the limit of large Euclidean time and therefore not suitable for our setup \cite{Philipsen1998,Bali2005,Rothe2006}. Instead, to detect strings and string breaking in the dynamical setup, we propose three different observables. We monitor their values throughout the evolution, taking advantage of the fact that we have access to the MPS wave function at all times.

In the first place, the spatially resolved spin and flux configurations in the system allow us to visualize the change with respect to the initial configuration (see e.g. figure ~\ref{fig:observables_timag_3D}).
 
A second observable is the local imbalance between red and green species, also spatially resolved. A string operator (\ref{string_op}) changes the fermionic content of the sites at its end-points, such that, on the strong coupling vacuum, it will produce a superposition of states having a single red or a single green fermion at the beginning or at the end of the string. This can be detected by the operator ${Q^\alpha_n}^2=\frac{1}{4}(n_{r,n}-n_{g,n})^2$, where $n_{r,n}$ and $n_{g,n}$ are the occupation numbers for the two fermionic species on site $n$.\footnote{One should note that the index $\alpha$ in ${Q^\alpha_n}^2$ is not summed. This might look puzzling at first, as it seems that we are using a non color-neutral object as observable. However, a little calculation shows (see appendix \ref{app:spin_formulation}) that ${Q^\alpha_n}^2$ is identical for all $\alpha$. Therefore summing $\alpha$ would only yield an additional factor of 3 which we are not taking into account.}
 
A third observable can be proposed that looks only at the flux content of the links. Applying a string of length $l$, starting at site $n$, on the strong coupling vacuum produces a state in which links between sites $n$ and $n+l$ carry non-vanishing flux, whereas outside the region there is no flux. We can construct projectors $P_{nl}$ on this kind of configurations. However, we are not interested in a single string but rather in the statistics of string lengths in a (time dependent) state $|\phi(t)\rangle$. Thus, we bin all strings of a certain length together and normalize by the number of possible strings of length $l$:
\begin{align*}
 P_l=\frac{\sum_{n=1}^{N-l}\langle\phi(t)|P_{nl}|\phi(t)\rangle}{N-l}.
\end{align*}
In this way we can obtain histograms for the distribution of the string lengths in the state at a given time $t$. 

As we show in the next section, these three observables allow us to reliably determine whether our system still contains the initially imposed string or the string is broken.

\section{\label{sec:static_qq}Ground state with static external charges}

The usual way of probing for string breaking in lattice Monte Carlo studies is the analysis of the static quark-antiquark potential. Using the MPS method, we can also determine the energy of an extra pair of external charges as a function of their distance by simply computing the ground state in a sector, where two external static charges are placed at the desired separation.
 
In particular, we choose two static external charges $q^y=\pm 1/2$, located in the central region of the system (to minimize finite-size effects) at a distance $l$. Gauge invariance requires that there is a flux tube connecting them. In the strong coupling vacuum the charge at each site is zero, as well as the left and right electric field on each link. Consequently, we can prepare a state with the desired external charge distribution by applying the (non-invariant) operator $U_n \dots U_{n+l-1}$ on the strong coupling ground state. This operator only acts on links from $n$ to $n+l-1$, on which it creates a finite electric field. Hence the Gauss Law at the beginning and the end of the string yields a non-zero value corresponding to a state with $q^y_n=\pm 1/2$ and $q^y_{n+l}=\mp 1/2$. Subsequently we evolve this state in imaginary time to determine the ground state of (\ref{KogutSusskindHamiltonian}) for a chosen set of parameters, $(\varepsilon,m,g)$. For the results presented in this section we use a time step $\Delta t=\num{1.0d-3}$ and a bond dimension $D=100$, parameters which turn out to be sufficient to avoid noticeable numerical errors (see appendix \ref{app:errors} for a more detailed error analysis).

\begin{figure}[!htbp]
\centering
\includegraphics[width=\columnwidth]{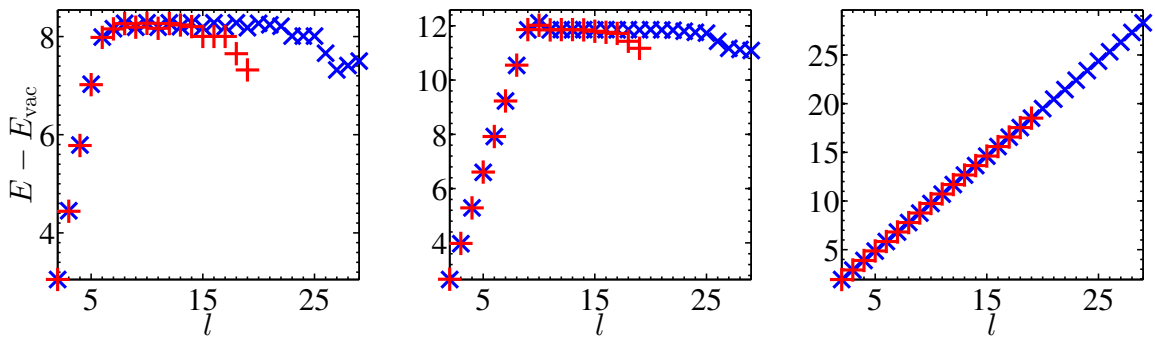}
\caption{Ground state energy for $N=22$ (red crosses) and $N=30$ (blue \X's ) as a function of the initial separation of the external charges for $m=3.0$ (left panel), $m=5.0$ (central panel) and $m=10.0$ (right panel).}
\label{fig:static_qq}
\end{figure}

In figure \ref{fig:static_qq} we compare the ground state energy $E$ for $\varepsilon=3.0$, $g=1.0$ and various masses, where we subtracted the energy of the interacting vacuum without external charges $E_\mathrm{vac}$. For $m=3.0$ and $m=5.0$ we observe three regions as functions of the string length. For short strings the energy grows linearly with the string length, indicating the stretching flux tube between the charges, namely the presence of the string in the ground state. From a certain value $l_c$ on, the ground state energy does not depend on $l$. This is the signature for string breaking, as after reaching the threshold for creating particle-antiparticle pairs, reducing the flux is energetically favorable, the string breaks and the energy is independent of the initial string length. Finally we observe a third region, when the string length is already close to the system size, and finite size effects become noticeable. 
In our plots we clearly see that the values of $l_c$, where one transitions from the string region to the breaking region, are independent of the system size. For $m=10.0$ the mass is large enough that one does not leave the linear scaling region, even if we create the longest string that fits in our system. Hence, we do not expect string breaking to occur in this system.

\begin{figure}[!htbp]
\centering
\includegraphics[width=\columnwidth]{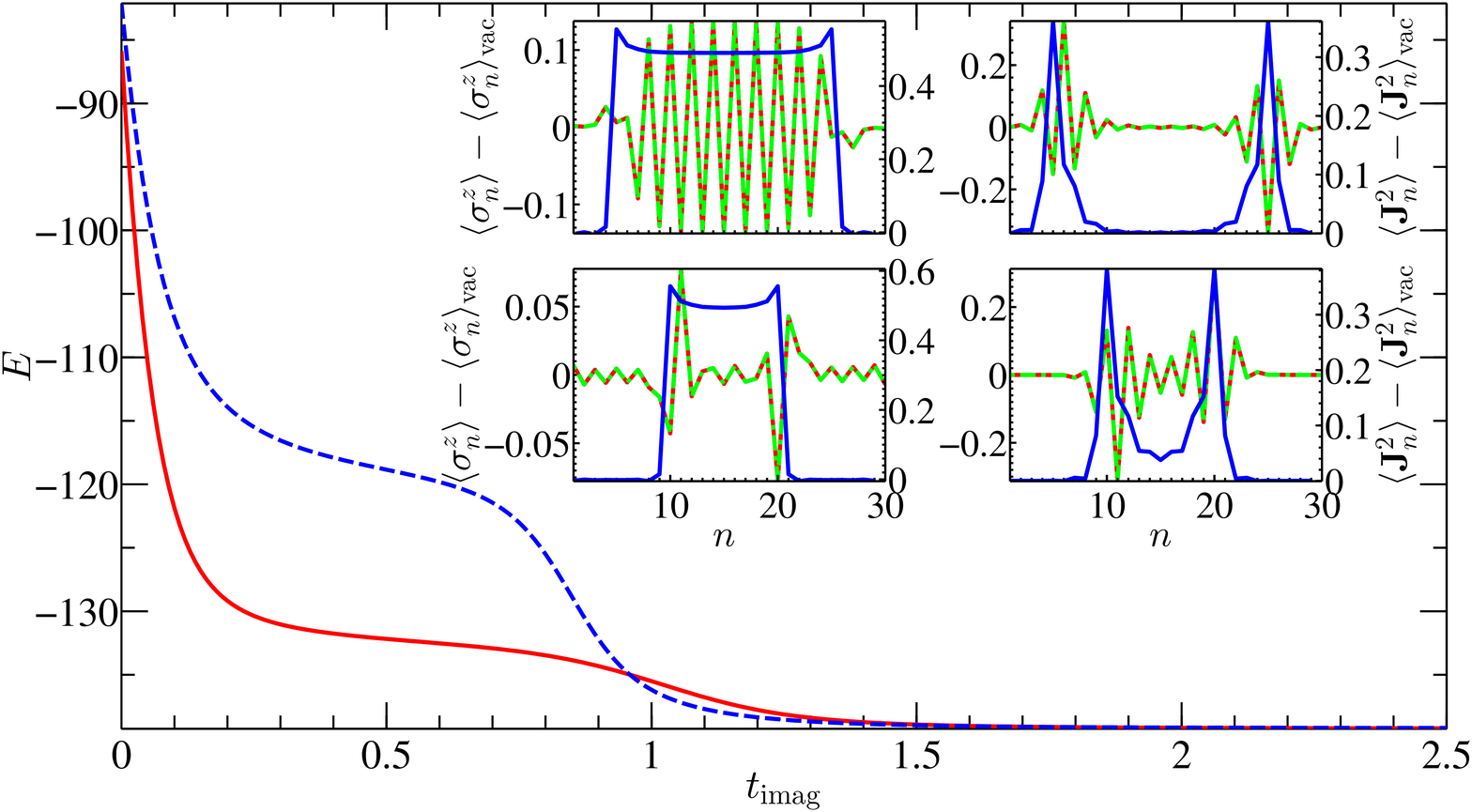}
\caption{Energy as function of imaginary time for $N=30$, $m=3.0$, $l=11$ (red solid line) and $l=21$ (blue dashed line). The insets show the difference of expectation values for $\sigma_{r,n}^z$ (red lines, left $y$-axes), $\sigma_{g,n}^z$ (green lines, left $y$-axes), and $\mathbf{J}^2_n$ (blue lines, right $y$-axes) with respect to the interacting vacuum, shortly before and after the drop in energy (left upper panel: $l=21$, $t=0.75$, right upper panel: $l=21$, $t=1.25$;  left lower panel: $l=11$, $t=0.5$, right lower panel: $l=11$, $t=1.25$).}
\label{fig:E_vs_timag}
\end{figure}

\begin{figure}[!htbp]
\centering
\includegraphics[width=\columnwidth]{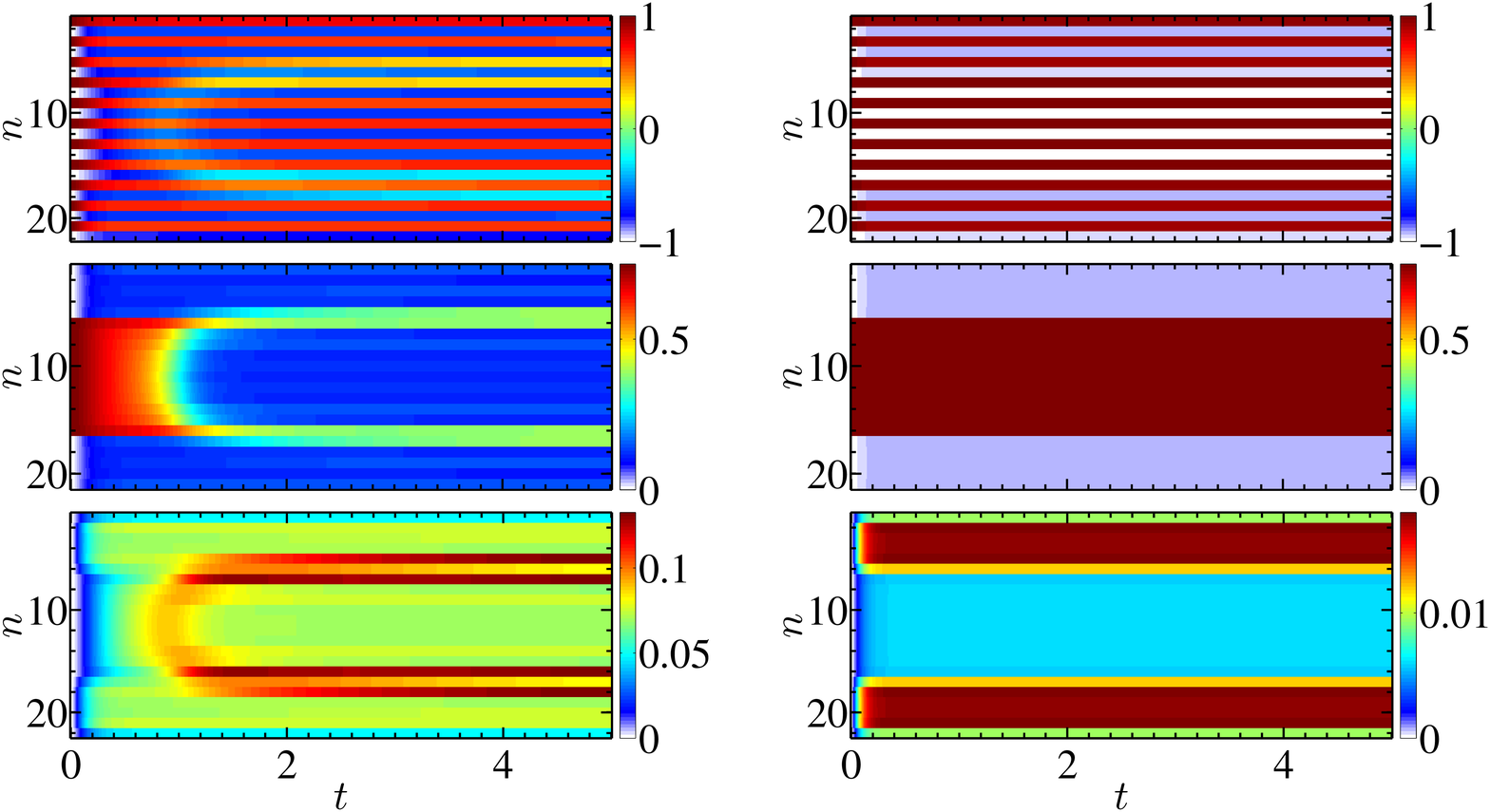}
\includegraphics[width=\columnwidth]{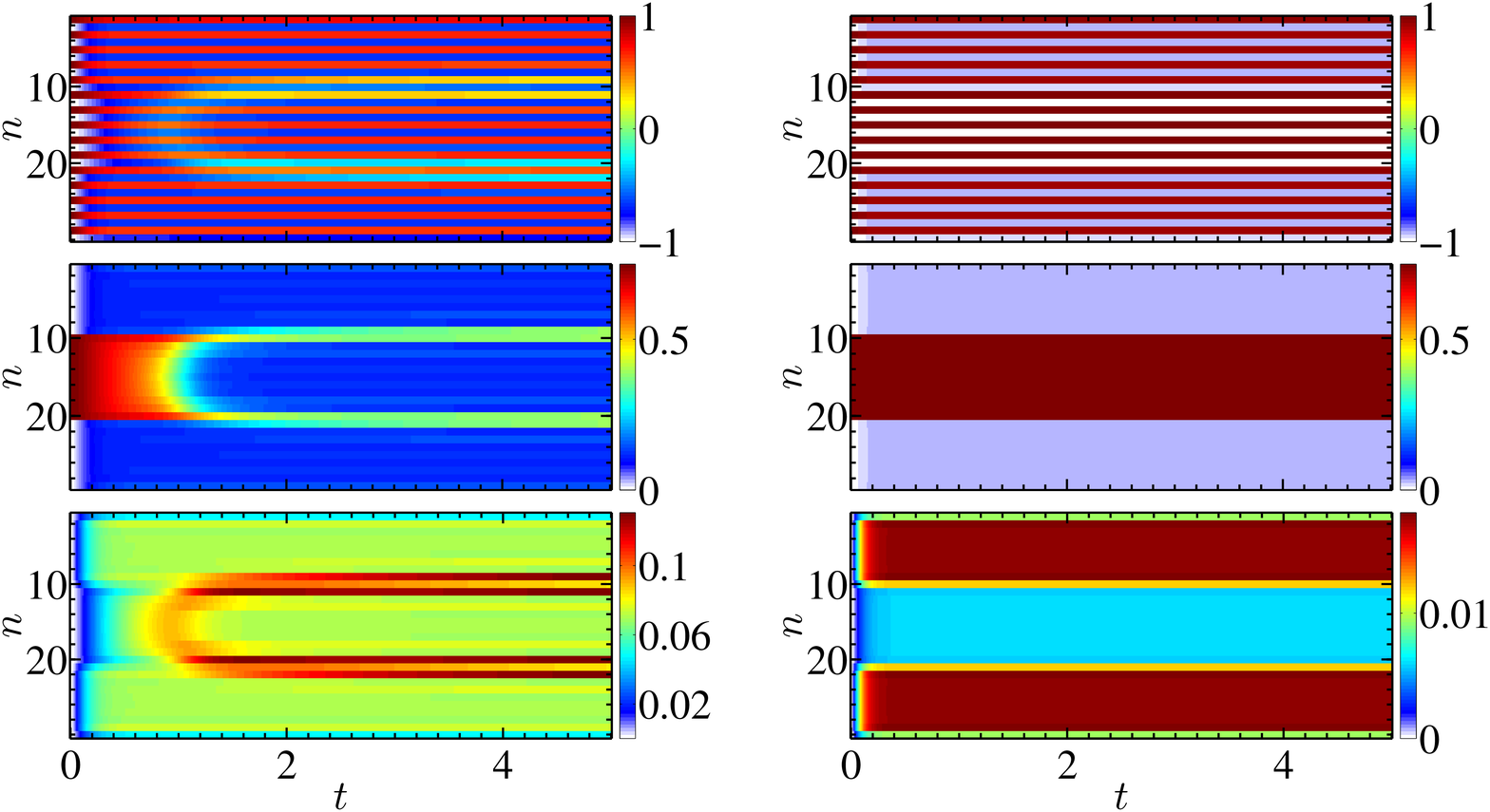}
\caption{Imaginary-time evolution of a string with length $l=11$ between two external static charges, in a system of size $N=22$ (upper three rows) and $N=30$ (lower three rows), for dynamical fermion masses $m=3.0$ (left column) and $10.0$ (right column). Shown are the site resolved expectation values for $\sigma^z_{r,n}$, $\sigma^z_{g,n}$ (first and fourth row), $\mathbf{J}^2_n$ (second and fifth row) and ${Q_n^\alpha}^2$ (third and sixth row) as a function of imaginary time.}
\label{fig:observables_timag_3D}
\end{figure}

The difficulty of numerically detecting string breaking with Monte Carlo techniques has been traced back to the mixing of two-meson states being hard to capture by Wilson loops \cite{Michael1992,Philipsen1998,Knechtli1998}. With our method we can explicitly see this mixing happening, as illustrated by figure \ref{fig:E_vs_timag}. The figure shows the energy as a function of imaginary time for several cases in the breaking region. At the beginning of the imaginary time evolution, we observe the energy going down until it reaches a metastable plateau, corresponding to the interacting string state, evidenced by the spin and flux configurations at that point, shown in the left inset panels. For later times there is then again a significant decrease in energy corresponding to the breaking of the string.  This can also be seen in the spin and flux configuration (right inset panels of figure \ref{fig:E_vs_timag}), where the region of high flux between the static charges is going away after the decrease in energy, and only a peak in the flux around the external charges remains. To better compare the string breaking scenario with that of a string ground state, the whole imaginary time evolution of spin, flux and charge square configurations for masses $m=3.0$ and $10.0$ and system sizes $N=22$ and $30$ is shown in figure \ref{fig:observables_timag_3D}. The flux configuration is giving an indication that the particle-antiparticle pairs created during the breaking process are clustering around the heavy external charges and screen the electric flux. Furthermore figure \ref{fig:observables_timag_3D} reveals that there is essentially no difference between both system sizes.

\begin{figure}[!htbp]
\centering
\includegraphics[width=\columnwidth]{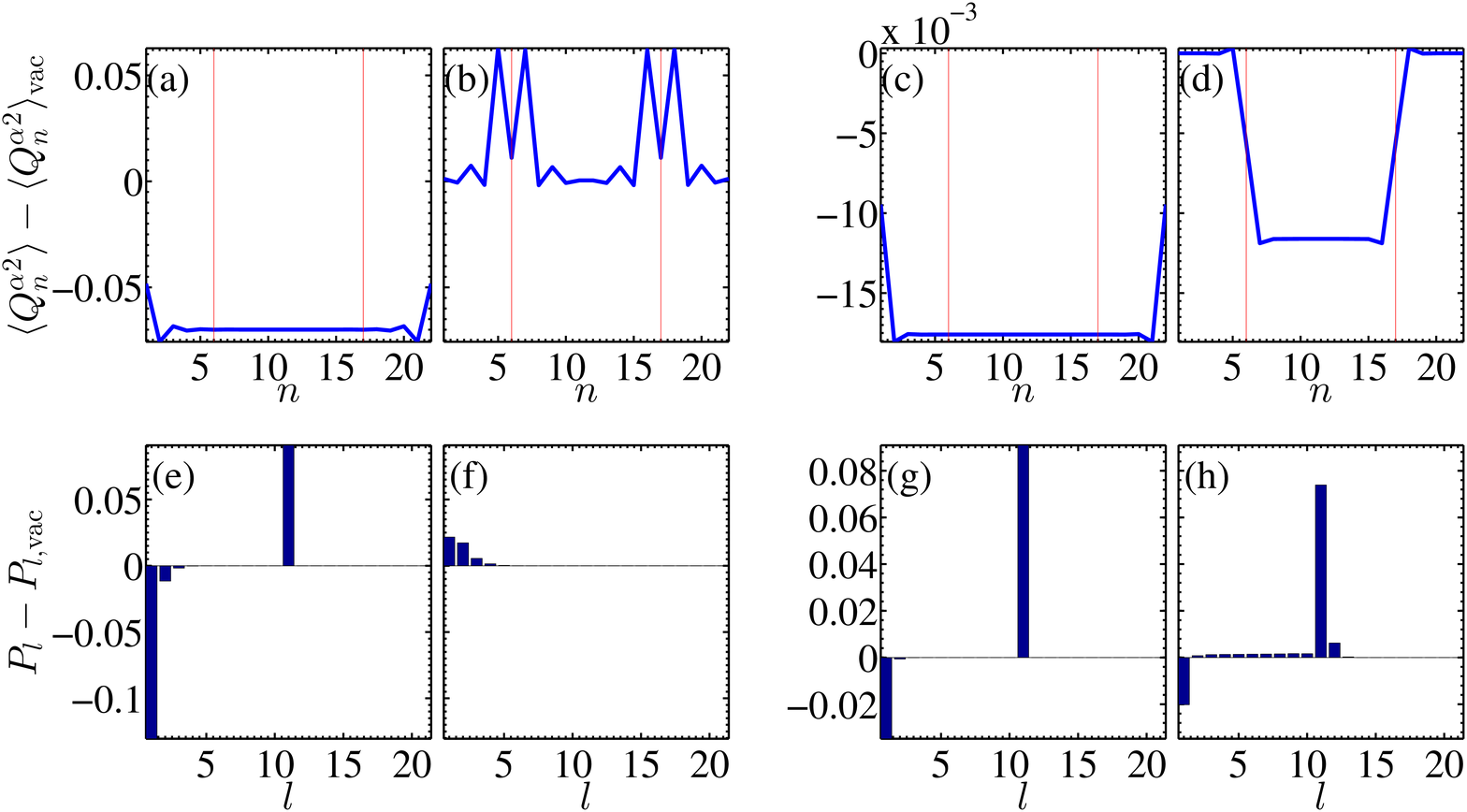}
\caption{Charge square configuration (upper panels) and histograms of string lengths (lower panels) at the beginning and the end of the evolution for a system of size $N=22$ and $l=11$, the vertical lines in the upper panels indicate the position of the external charges. Panel (a) and (e) show the data for $m=3.0$ at the beginning, panel (b) and (f) for $m=3.0$ at the end, panel (c) and (g) for $m=10.0$ at the beginning and panel (d) and (h) for $m=10.0$ at the end of the evolution.}
\label{fig:observables_timag}
\end{figure}

In figure \ref{fig:observables_timag} we plot the charge squared ${Q^\alpha_n}^2$ and the histograms $P_l$ for the initial configuration and for the final ground state respectively. We subtract the configuration of the interacting vacuum in the sector without external charges, in order to better visualize the difference. As one can see, in the breaking case two peaks around the external charges form in the charge square configuration, thus verifying that the particles created during the breaking process indeed cluster around the external charges and screen the flux. By contrast, in the nonbreaking case the charge square configuration only changes slightly. The histograms for the string lengths show a similar picture. In the nonbreaking case the clear initial peak at $l=11$ is preserved, whereas for $m=3.0$ it vanishes and peaks emerge  around smaller string length.

From figure \ref{fig:static_qq} we can identify the parameter regions in which we expect string breaking to occur. To study the real-time dynamics of the string breaking we select two distinct situations. We choose $l=11$, which is deeply in the breaking region for $m=3.0$ but still far enough from the point where finite size effects are noticeable, and for which we do not expect any string breaking with $m=10.0$. 
For all the following real-time cases we set $\varepsilon=3.0$ and $g=1.0$, as in the imaginary-time setup.

\section{\label{sec:dynamic_qq}Real-time evolution with static external charges}
\begin{figure}[!htbp]
\centering
\includegraphics[width=\columnwidth]{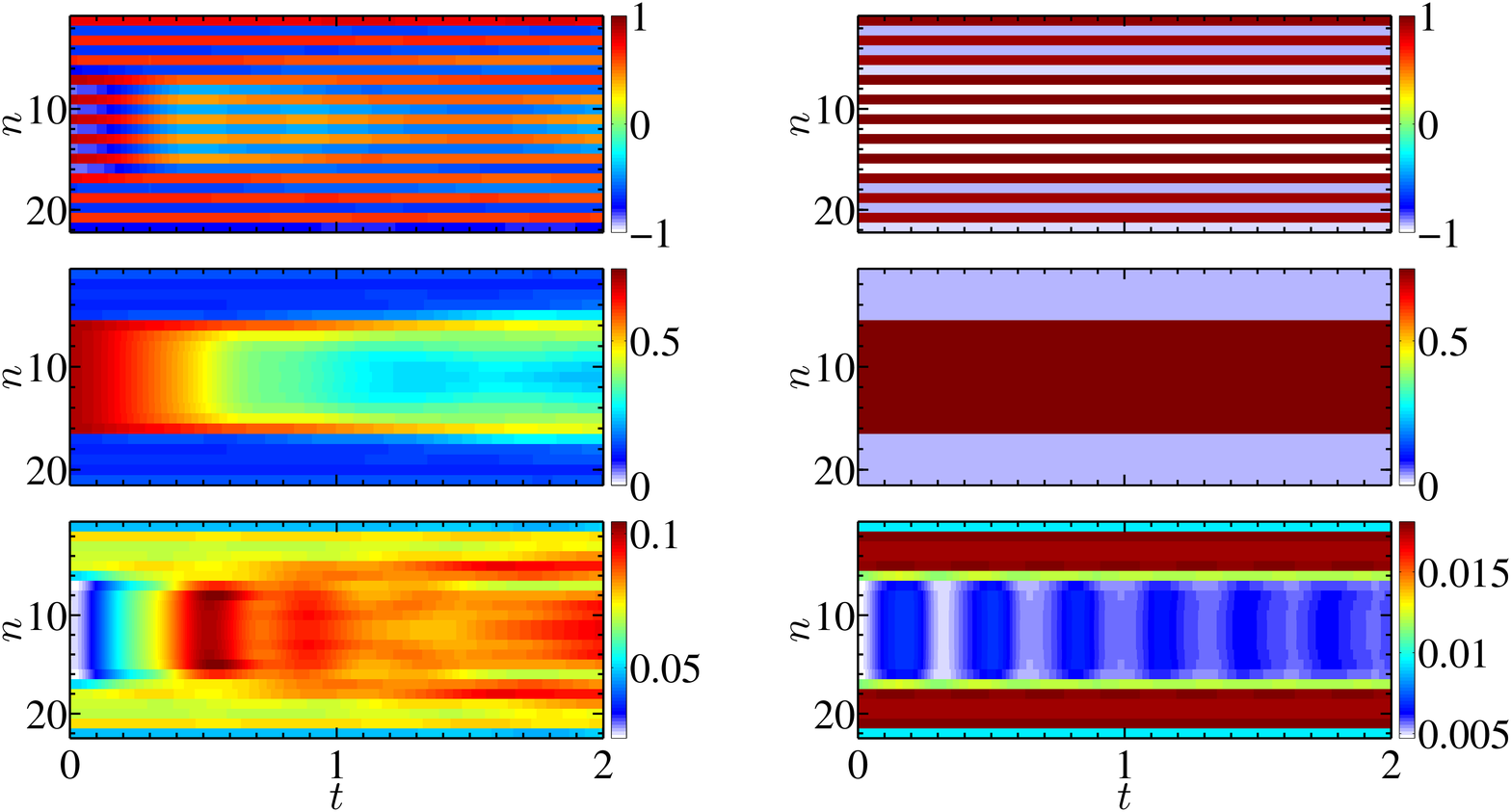}
\includegraphics[width=\columnwidth]{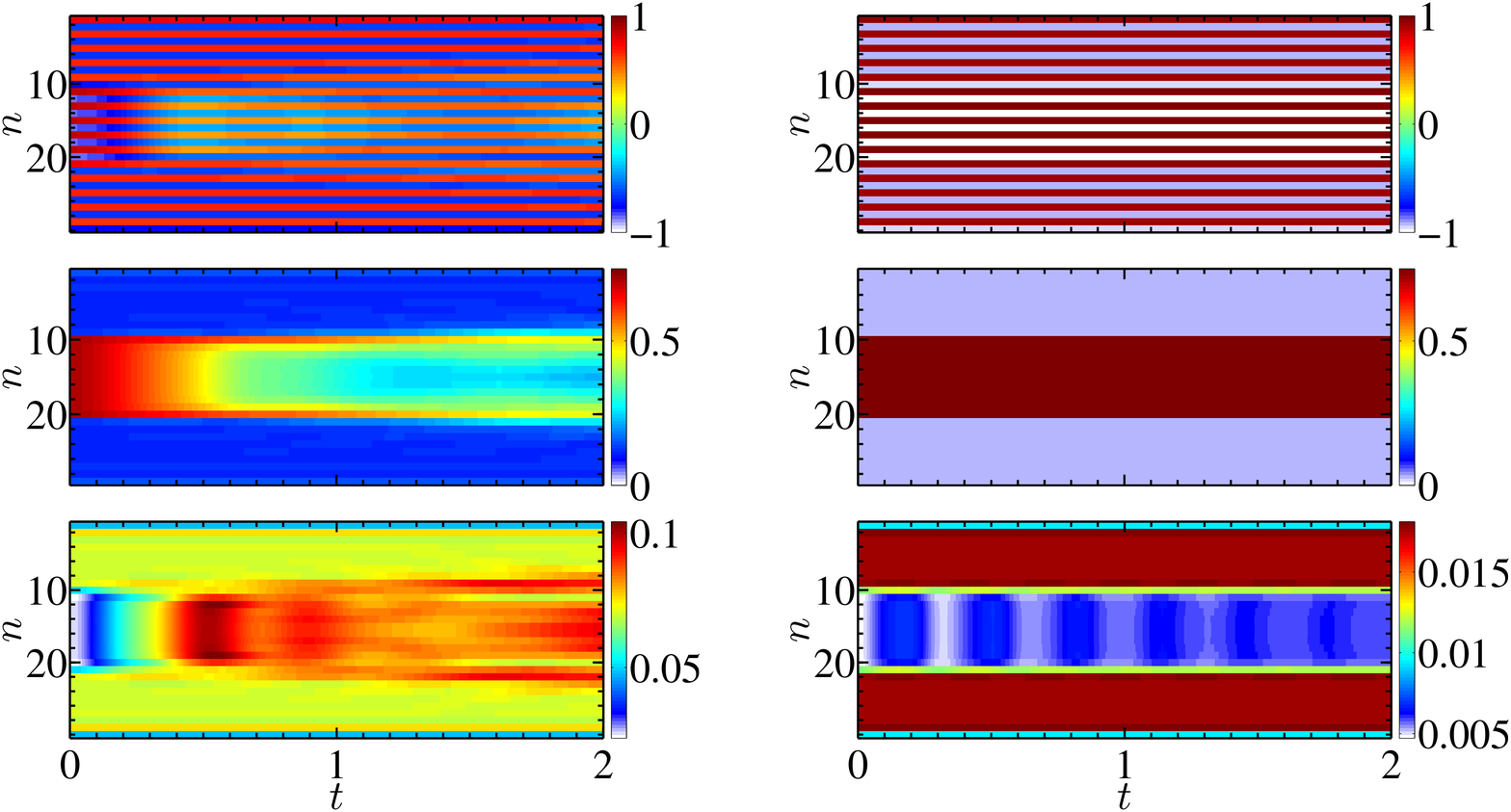}
\caption{Real-time evolution of a string with length $l=11$ between two external static charges, in a system of size $N=22$ (upper three rows) and $N=30$ (lower three rows), for dynamical fermion masses $m=3.0$ (left column) and $10.0$ (right column). Shown are the site resolved expectation values for $\sigma^z_{r,n}$, $\sigma^z_{g,n}$ (first and fourth row), $\mathbf{J}^2_n$ (second and fifth row) and ${Q_n^\alpha}^2$ (third and sixth row) as a function of time.}
\label{fig:observables_real_ext_3D}
\end{figure}

\begin{figure}[!htbp]
\centering
\includegraphics[width=\columnwidth]{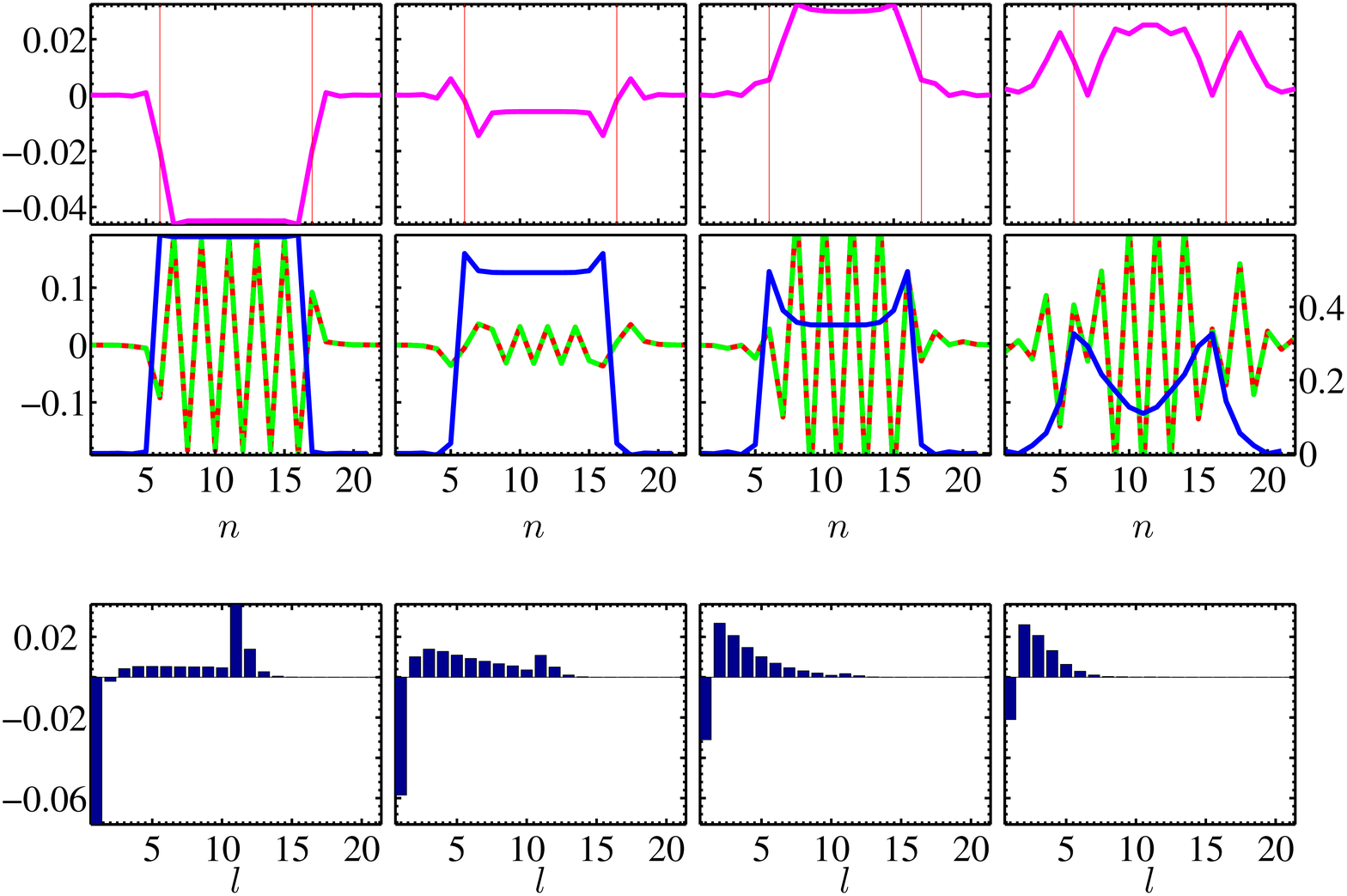}
\caption{Real-time snapshots for a system with external static charges at a distance $l=11$, dynamical fermion mass $m=3.0$ and size $N=22$. The upper row shows the charge square configuration, $\langle {Q^\alpha_n}^2\rangle - \langle {Q^\alpha_n}^2\rangle_\mathrm{vac}$, the central row the spin, $\langle\sigma_{r/g,n}^z\rangle - \langle\sigma_{r/g,n}^z\rangle_\mathrm{vac}$ (red and green lines, left $y$-axes), and flux configuration, $\langle \mathbf{J}^2_n\rangle - \langle \mathbf{J}^2_n\rangle_\mathrm{vac}$ (blue lines, right $y$-axes), and the lower row the histograms for the string lengths, $P_l - P_{l,\mathrm{vac}}$. The vertical red lines in the upper row indicate the position of the external charges. Each column corresponds to a time instant, $t=0$, $0.25$, $0.5$ and $2.0$.}
\label{fig:qq_m3}
\end{figure}

\begin{figure}[!htbp]
\centering
\includegraphics[width=\columnwidth]{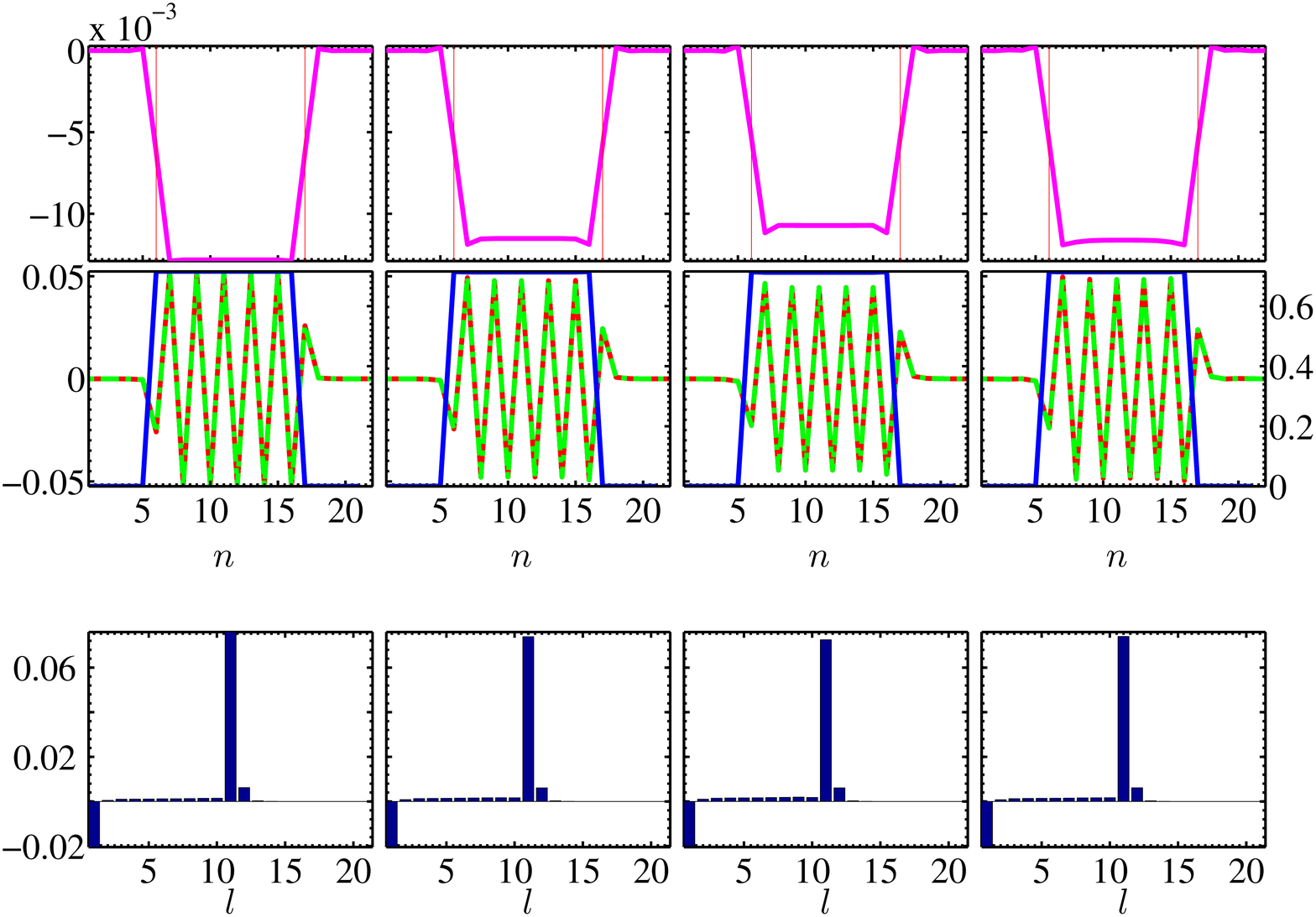}
\caption{Real-time snapshots for a system with external static charges at a distance $l=11$, dynamical fermion mass $m=10.0$ and size $N=22$. The upper row shows the charge square configuration, $\langle {Q^\alpha_n}^2\rangle - \langle {Q^\alpha_n}^2\rangle_\mathrm{vac}$, the central row the spin, $\langle\sigma_{r/g,n}^z\rangle - \langle\sigma_{r/g,n}^z\rangle_\mathrm{vac}$ (red and green lines, left $y$-axes), and flux configuration, $\langle \mathbf{J}^2_n\rangle - \langle \mathbf{J}^2_n\rangle_\mathrm{vac}$ (blue lines, right $y$-axes), and the lower row the histograms for the string lengths, $P_l - P_{l,\mathrm{vac}}$. The vertical red lines in the upper row indicate the position of the external charges. Each column corresponds to a time instant, $t=0$, $0.25$, $0.5$ and $2.0$.}
\label{fig:qq_m10}
\end{figure}

MPS methods give us access not only to the static properties, but also to the real-time dynamics of the system. We may then investigate how the string breaking process manifests dynamically when the static charges are introduced in the interacting vacuum. To this end, we first compute a MPS approximation to the interacting vacuum of the theory using variational energy minimization \cite{Verstraete2004,White1992,Schollwoeck2011}. Starting from this state, we apply again the non gauge invariant operator $U_n \dots U_{n+l-1}$, which effectively creates two static external charges $q^y=\pm 1/2$ separated by a distance $l=11$, in a gauge invariant manner, thus starting the connecting flux tube. Subsequently we evolve this state in real time with $\Delta t=\num{1.0d-4}$, $D=100$ and determine the time dependent spin, flux and charge square configurations along the chain, as shown in figure \ref{fig:observables_real_ext_3D}. As in the imaginary time case, one can see that a system size $N=22$ is sufficient to avoid finite size effects. In order to better appreciate the dynamics, we show the details of the three proposed observables at fixed times in the evolution, $t=0$, $0.25$, $0.5$ and $2$, for each mass in figures \ref{fig:qq_m3} and \ref{fig:qq_m10}, where we again subtracted the interacting vacuum configuration to visualize the difference to the ground state without external charges.

For $m=3.0$ the plots in figure \ref{fig:qq_m3} reveal that between $t=0$ and $t=0.25$ particle-antiparticle pairs are created inside the string region, leading to a significant increase in ${Q^\alpha_n}^2$ there. This is accompanied with a considerable decrease of the initial peak in the string length histogram around $l=11$ and a change in the spin and flux configurations. This growth is continuing up to $t=0.5$ where there has been a large amount of particles created in the string region and the peak in the histogram at $l=11$ is already gone. At later times, $t=2.0$, these particles are clustering around the region of the external charges and, as the flux configuration reveals, the external charges are screened, leading to a reduction of flux in the center region of the original string. For $m=10.0$ the picture is significantly different, as there is essentially no change during the evolution in either the charge, spin and flux configurations or the string length histograms, which show a single dominant peak at $l=11$ at all times, therefore indicating that the initial string is preserved during the evolution. The minor changes over time present in figure \ref{fig:qq_m10} result from the fact that the starting state of the evolution is not an eigenstate of the Hamiltonian, and consequently it is not perfectly steady.

\section{\label{sec:dynamic_string}Real-time evolution with dynamical charges}

The scenarios with external static charges allow us to isolate and study the string breaking phenomenon. However, it is also worth investigating the alternative scenario in which the charges added to the vacuum are themselves dynamical, as this is arguably a more realistic situation. This enables additional configurations, in which the charges can move, to also play a role in the evolution, so that the string breaking phenomenon may be displayed differently.

\begin{figure}[!htbp]
\centering
\includegraphics[width=\columnwidth]{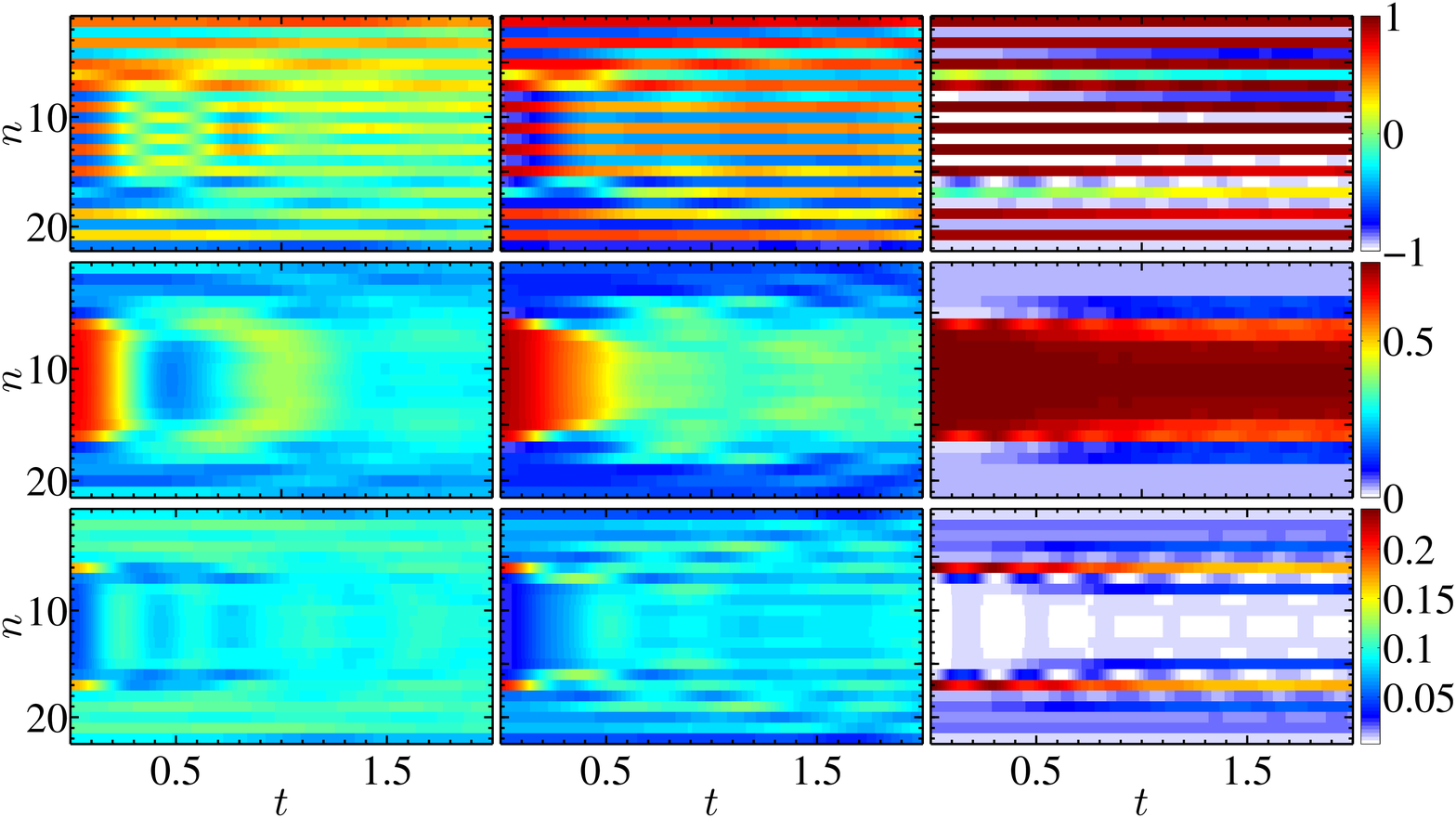}
\includegraphics[width=\columnwidth]{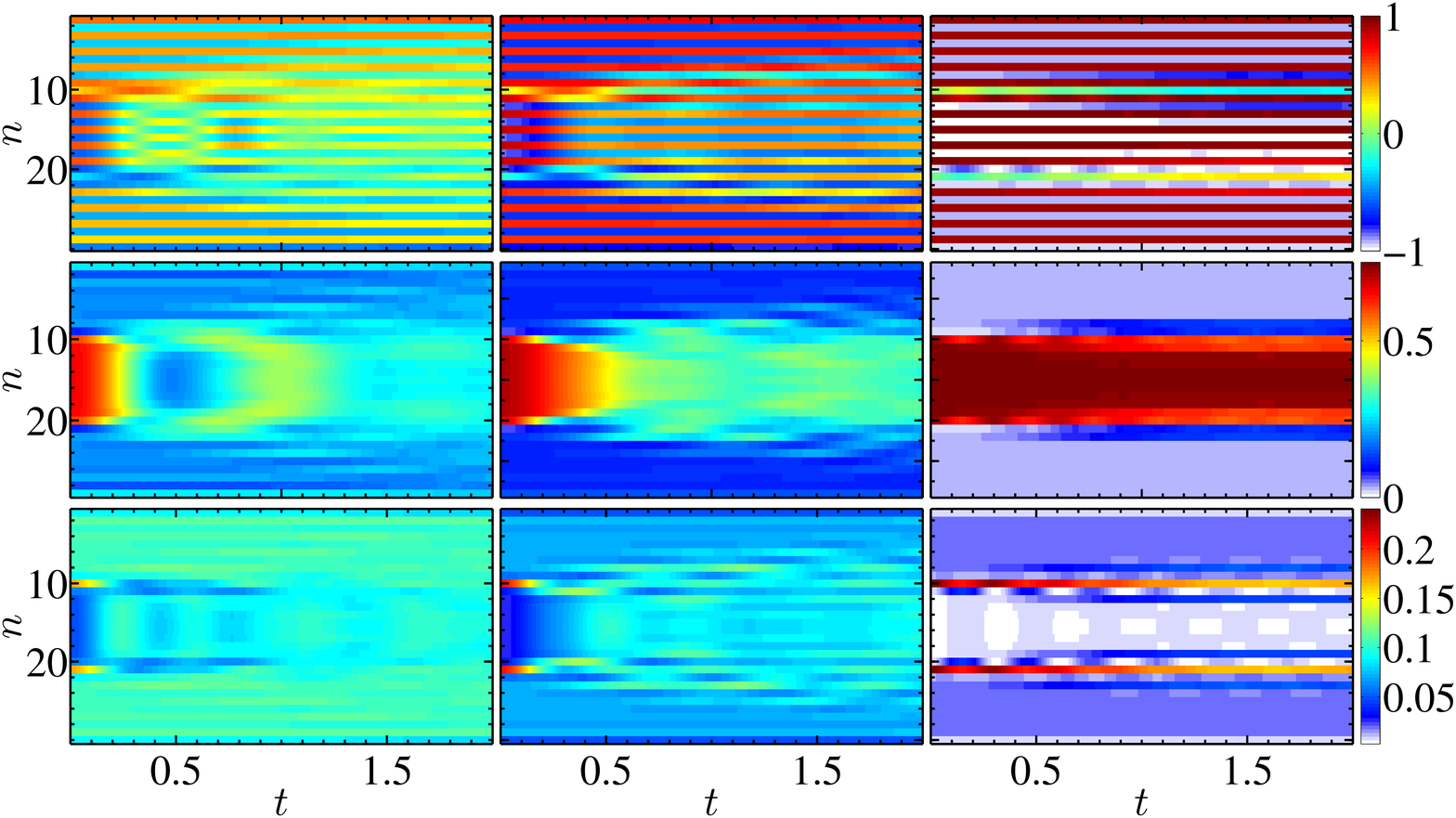}
\caption{Real-time evolution of a string with length $l=11$ between fully dynamical fermions, in a system of size $N=22$ (upper three rows) and $N=30$ (lower three rows), for dynamical fermion masses $m=1.0$ (left column), $m=3.0$ (central column) and $10.0$ (right column). Shown are the site resolved expectation values for $\sigma^z_{r,n}$, $\sigma^z_{g,n}$ (first and fourth row), $\mathbf{J}^2_n$ (second and fifth row) and ${Q_n^\alpha}^2$ (third and sixth row) as a function of time.}
\label{fig:observables_real_3D}
\end{figure}

\begin{figure}[!htbp]
\centering
\includegraphics[width=\columnwidth]{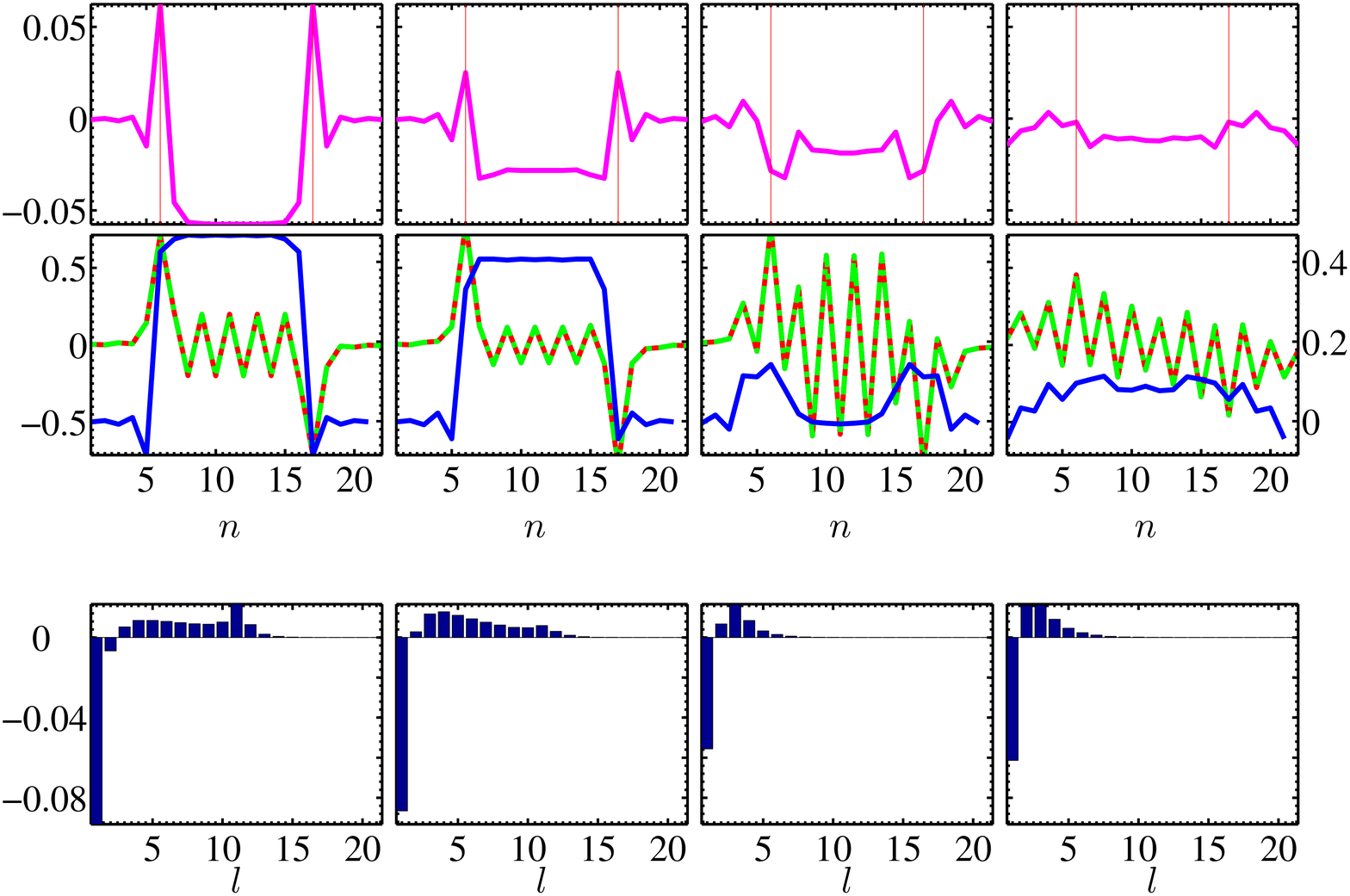}
\caption{Real-time snapshots from the evolution of a string with length $l=11$ between fully dynamical fermions, for $m=1.0$ and system size $N=22$. The upper row shows the charge square configuration, $\langle {Q^\alpha_n}^2\rangle - \langle {Q^\alpha_n}^2\rangle_\mathrm{vac}$, the central row the spin, $\langle\sigma_{r/g,n}^z\rangle - \langle\sigma_{r/g,n}^z\rangle_\mathrm{vac}$ (red and green lines, left $y$-axes), and flux configuration, $\langle \mathbf{J}^2_n\rangle - \langle \mathbf{J}^2_n\rangle_\mathrm{vac}$ (blue lines, right $y$-axes), and the lower row the histograms for the string lengths, $P_l - P_{l,\mathrm{vac}}$. The vertical red lines in the upper row indicate the initial position of the charges added to the vacuum. Each column corresponds to a time instant, $t=0$, $0.125$, $0.5$ and $1.5$. }
\label{fig:string_m1}
\end{figure}

\begin{figure}[!htbp]
\centering
\includegraphics[width=\columnwidth]{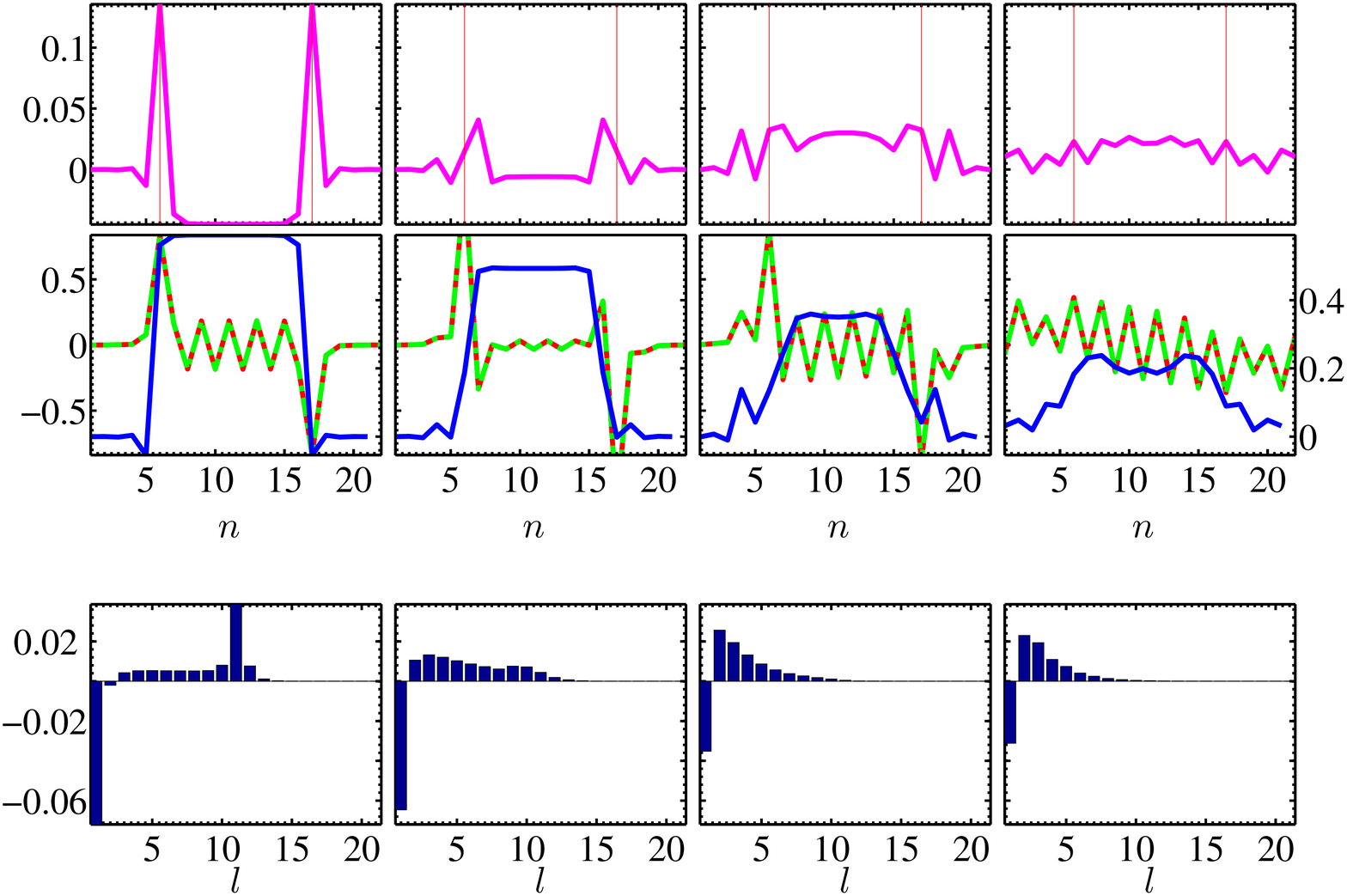}
\caption{Real-time snapshots from the evolution of a string with length $l=11$ between fully dynamical fermions, for $m=3.0$ and system size $N=22$. The upper row shows the charge square configuration, $\langle {Q^\alpha_n}^2\rangle - \langle {Q^\alpha_n}^2\rangle_\mathrm{vac}$, the central row the spin, $\langle\sigma_{r/g,n}^z\rangle - \langle\sigma_{r/g,n}^z\rangle_\mathrm{vac}$ (red and green lines, left $y$-axes), and flux configuration, $\langle \mathbf{J}^2_n\rangle - \langle \mathbf{J}^2_n\rangle_\mathrm{vac}$ (blue lines, right $y$-axes), and the lower row the histograms for the string lengths, $P_l - P_{l,\mathrm{vac}}$. The vertical red lines in the upper row indicate the initial position of the charges added to the vacuum. Each column corresponds to a time instant, $t=0$, $0.25$, $0.5$ and $2.0$.}
\label{fig:string_m3}
\end{figure}

\begin{figure}[!htbp]
\centering
\includegraphics[width=\columnwidth]{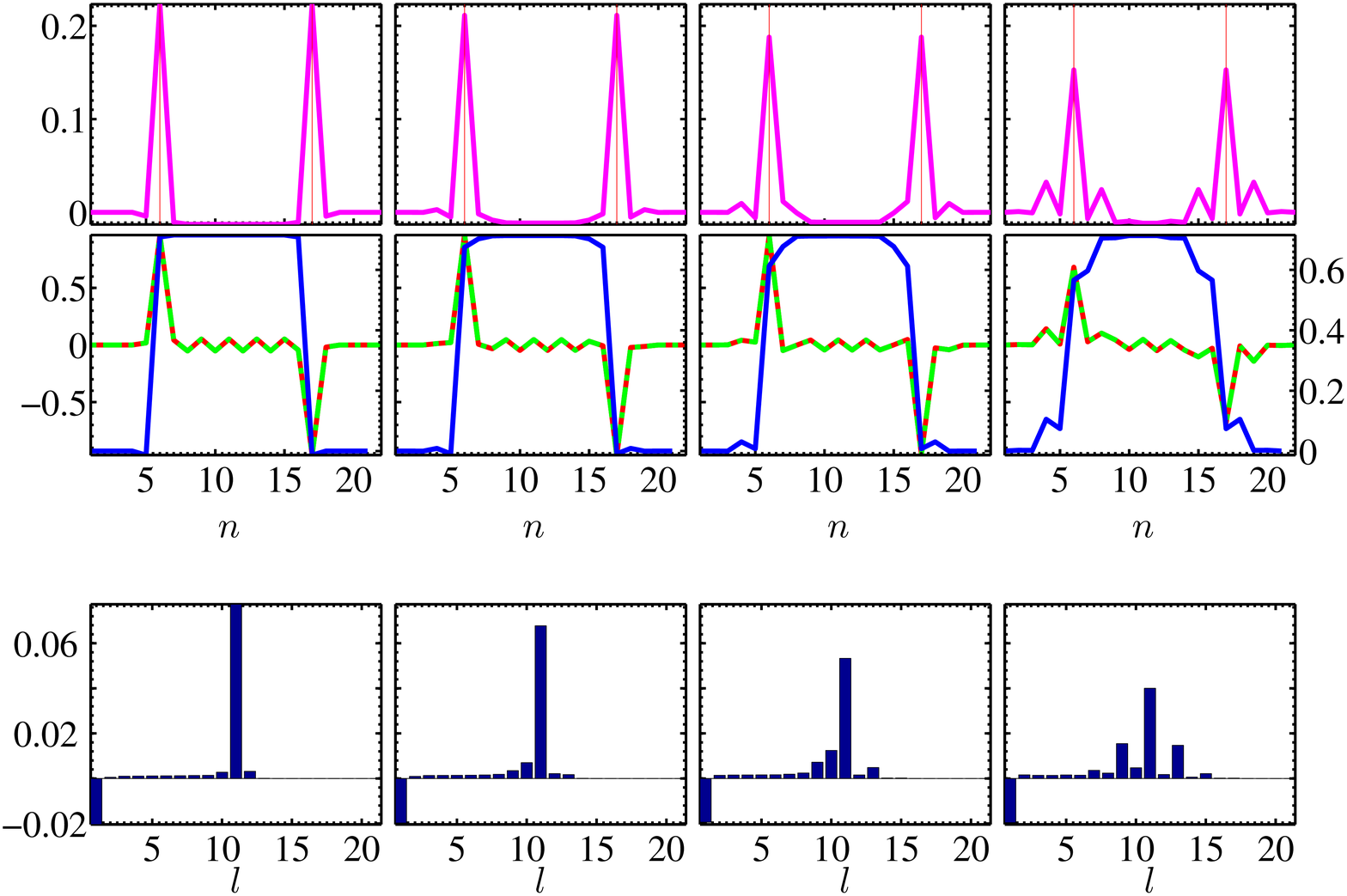}
\caption{Real-time snapshots from the evolution of a string with length $l=11$ between fully dynamical fermions, for $m=10.0$ and system size $N=22$. The upper row shows the charge square configuration, $\langle {Q^\alpha_n}^2\rangle - \langle {Q^\alpha_n}^2\rangle_\mathrm{vac}$, the central row the spin, $\langle\sigma_{r/g,n}^z\rangle - \langle\sigma_{r/g,n}^z\rangle_\mathrm{vac}$ (red and green lines, left $y$-axes), and flux configuration, $\langle \mathbf{J}^2_n\rangle - \langle \mathbf{J}^2_n\rangle_\mathrm{vac}$ (blue lines, right $y$-axes), and the lower row the histograms for the string lengths, $P_l - P_{l,\mathrm{vac}}$. The vertical red lines in the upper row indicate the initial position of the charges added to the vacuum. Each column corresponds to a time instant, $t=0$, $0.25$, $0.5$ and $2.0$.}
\label{fig:string_m10}
\end{figure}

Using the same MPS techniques, we can also explore this setup. Thus we repeat the simulations described in section \ref{sec:dynamic_qq} but applying the gauge invariant string operator from eq. (\ref{string_op}) on the interacting vacuum to construct our initial state. This again results in a state with a string between two charges. Different to the systems studied earlier, the charges are not external, but they are now created on a site and are fully dynamical. Again, we study the time evolution of the spin, flux and charge square configurations for different fermion masses and system sizes, as shown in figure \ref{fig:observables_real_3D}, where we use $\Delta t=\num{1.0d-4}$ and $D=100$ as in the previous section. We compare the case $m=10.0$, in which the string does not break (see also figure \ref{fig:string_m10}), with $m=3.0$. In the latter case, which exhibited clear breaking for static external charges, the fully dynamical situation shows differences (compare figures \ref{fig:qq_m3} and \ref{fig:string_m3}). To better identify the features of string breaking, we look at an even smaller fermion mass, $m=1.0$, shown in figure \ref{fig:string_m1}.

In all cases we see a clear initial peak in the charge square configuration at the beginning and at the end of the string. For $m=1.0$ and $3.0$ new charges emerge especially in the string region and these peaks are quickly decaying. Also the flux in the string region is decaying while two small peaks are preserved roughly around the start and end point of the original string. The histograms for the distribution of string lengths show a similar picture: at the beginning of the evolution there is a clear peak at around $l=11$ which is gone at around $t=0.25$. In the $m=1.0$ case the initial peak at $l=11$ is less dominant as in the $m=3.0$ case, due to the fact that in this case the interaction strength $\varepsilon$ is more important, leading to a state less close to a strong coupling string. For later times one sees that in both cases smaller string lengths are dominating in the system, therefore indicating that the string is broken. By contrast, in the case of $m=10.0$ the dominant peak at $l=11$ is preserved during the entire evolution. Although the magnitude decreases, one can see from the flux and charge square configurations that there is not a lot of change, which indicates that the original string is still present. The slight changes have the same origin as in the previous cases. As we are not starting with an eigenstate of the Hamiltonian but rather do a local quench, the state is not perfectly steady. Furthermore in this case the charges are fully dynamical, what leads to richer dynamics.

\section{Conclusion}
\label{sec:conclusions}
We have studied the real-time dynamics of string breaking with MPS in a truncated 1+1 dimensional SU(2) lattice gauge theory. By looking at the charge square, the flux configuration and the statistics of string lengths, we are able to detect and characterize string breaking in statical and dynamical scenarios.

We calculate the ground state in the sector with two static external charges. This allows us to clearly pick up the static signature of string breaking, and to identify parameter regions where we expect string breaking to occur. Furthermore our calculation shows that the string state is metastable during the imaginary time evolution, until configurations with a broken string mix in and reduce the energy, as expected. We have also shown that the particle-antiparticle pairs resulting from the breaking string cluster around the external charges and subsequently screen the electric field. This can be detected with the spin and flux configuration as well as in the charge square and the distribution of string lengths in the system.

Using real-time simulations with MPS, we have studied the dynamics of string breaking in a setup with two static external charges. With the observables proposed before, the breaking and non-breaking cases are clearly distinguishable. We can explicitly observe that, in case the string is breaking, dynamical fermions are created that cluster around the external charges and screen them, therefore reducing the flux in the system. 

Finally, we have also simulated the time evolution of a string between fully dynamical fermions. We also identify situations in which the string breaks in this case. In particular, the decay of charges and the reduction of the flux in the middle of the string region indicate string breaking. Due to the fact that the initially created particle-antiparticle pair is now dynamical, we do not observe the clustering of the dynamically created charges, but rather a distribution along the system.

In our study we have used a 1+1 dimensional SU(2) gauge model. This model could be a suitable candidate for the implementation of a quantum simulator for a SU(2) lattice gauge theory. The observables and setups that we describe would also be viable for such a potential experimental realization. Furthermore the applied techniques are not limited to the specific model studied here. The truncation method used from Ref. \cite{Zohar2015} works for arbitrary compact (or finite) Lie groups and is not limited to the case of one spatial dimension. Therefore also other gauge groups could be studied and with more general TN such as PEPS \cite{Verstraete2004b} this study could also be extended to higher dimensions.

While completing the manuscript we became aware of a related work on TN and string breaking \cite{Pichler2015}.

\acknowledgments
We thank K. Jansen, B. Reznik, V. Vento and V. Gim\'enez for helpful discussions. This work was partially funded by EU through SIQS grant (FP7 600645). EZ acknowledges the support of the Alexander-von-Humboldt Foundation.

\bibliographystyle{JHEP}
\bibliography{Papers_MPQ_converted}
\appendix
\section{\label{app:spin_formulation}Spin formulation}
For the numerical implementation the fermionic degrees of freedom can be mapped to spins using a Jordan-Wigner transformation. Here we use the particular transformation from Ref. \cite{Steinhardt1977}:
\begin{align}
\begin{aligned}
\psi_{r,n} &= i\left[\prod_{l<n}i\sigma_{r,l}^zi\sigma_{g,l}^z\right]\sigma_{r,n}^-,\\
\psi_{g,n} &= -\left[\prod_{l<n}i\sigma_{r,l}^zi\sigma_{g,l}^z\right]\sigma_{r,n}^z\sigma_{g,n}^-.
\end{aligned}
\label{jordan_wigner_trafo}
\end{align}
The $\sigma$-matrices are the usual Pauli matrices, where the subscript indicates on which color on site $n$ they are acting. This labeling is convenient for practical purposes and completely equivalent to ordering the fermions as $\psi_{r,1},\psi_{g,1},\psi_{r,2},\psi_{g,2}\dots$ and doing a usual Jordan-Wigner transformation on a single fermionic field $\tilde{\psi}_n$ which is related to original ones as $\tilde{\psi}_1=\psi_{r,1}$, $\tilde{\psi}_2=\psi_{g,1}$, $\tilde{\psi}_3=\psi_{r,2}$, $\dots$. 

Using the transformation from eq. (\ref{jordan_wigner_trafo}), the Hamiltonian can be expressed in spin language as follows:
\begin{align*}
\begin{alignedat}{3}
&H = &&\varepsilon \sum_n\Bigl[\Bigr.&&\sigma^+_{r,n}\sigma^z_{g,n} U^{00}_{n}\sigma^-_{r,n+1} \\
                               & &&  &&+  i\sigma^+_{r,n}\sigma^z_{g,n} U^{01}_{n}\sigma^z_{r,n+1}\sigma^-_{g,n+1}\\
                               & &&  &&-i \sigma^+_{g,n} U^{10}_{n}\sigma^-_{r,n+1}\\
                               & &&  && +\sigma^+_{g,n} U^{11}_{n}\sigma^z_{r,n+1}\sigma^-_{g,n+1}\\
       & &&+ && \Bigl.h.c.\Bigr]\\
    & &&+&&\frac{m}{2}\sum_n (-1)^n\bigl[ (\sigma^z_{r,n}+1)+(\sigma^z_{g,n}+1)\bigr] \\
    &  &&+&&\frac{g^2}{2} \sum_n \mathbf{J}_{n}^2.
\end{alignedat}
\end{align*}
The charge operators $Q^\alpha_n=\frac{1}{2}\psi_n^\dagger\sigma^\alpha\psi_n$ can then be expressed in spin language as
\begin{align*}
\begin{aligned}
 Q^x_n&=\frac{1}{2}\left(-i\sigma^+_{r,n}\sigma^-_{g,n}+i\sigma^-_{r,n}\sigma^+_{g,n}\right),\\
 Q^y_n&=-\frac{1}{2}\left(\sigma^+_{r,n}\sigma^-_{g,n}+\sigma^-_{r,n}\sigma^+_{g,n}\right),\\
 Q^z_n&=\frac{1}{4}\left(\sigma^z_{r,n} - \sigma^z_{g,n}\right).
\end{aligned}
\end{align*}
A simple calculation yields for the charge square components
\begin{align*}
 {Q^\alpha_n}^2 = \frac{1}{8}\left(1-\sigma^z_{r,n}\sigma^z_{g,n}\right) \quad\quad \alpha=x,y,z.
\end{align*}

\section{\label{app:errors}Analysis of the numerical errors}
In this appendix we analyze the numerical errors of the results presented in sections \ref{sec:static_qq} - \ref{sec:dynamic_string}. In the main text we have already presented data for different system sizes thus showing that $N=22$ is enough to avoid noticeable finite size effects for the considered range of parameters. Here we focus on the influence of the time-step size and the bond dimension on our results.

\subsection{Ground state with static external charges}
\begin{figure}[!htbp]
\centering
\includegraphics[width=\columnwidth]{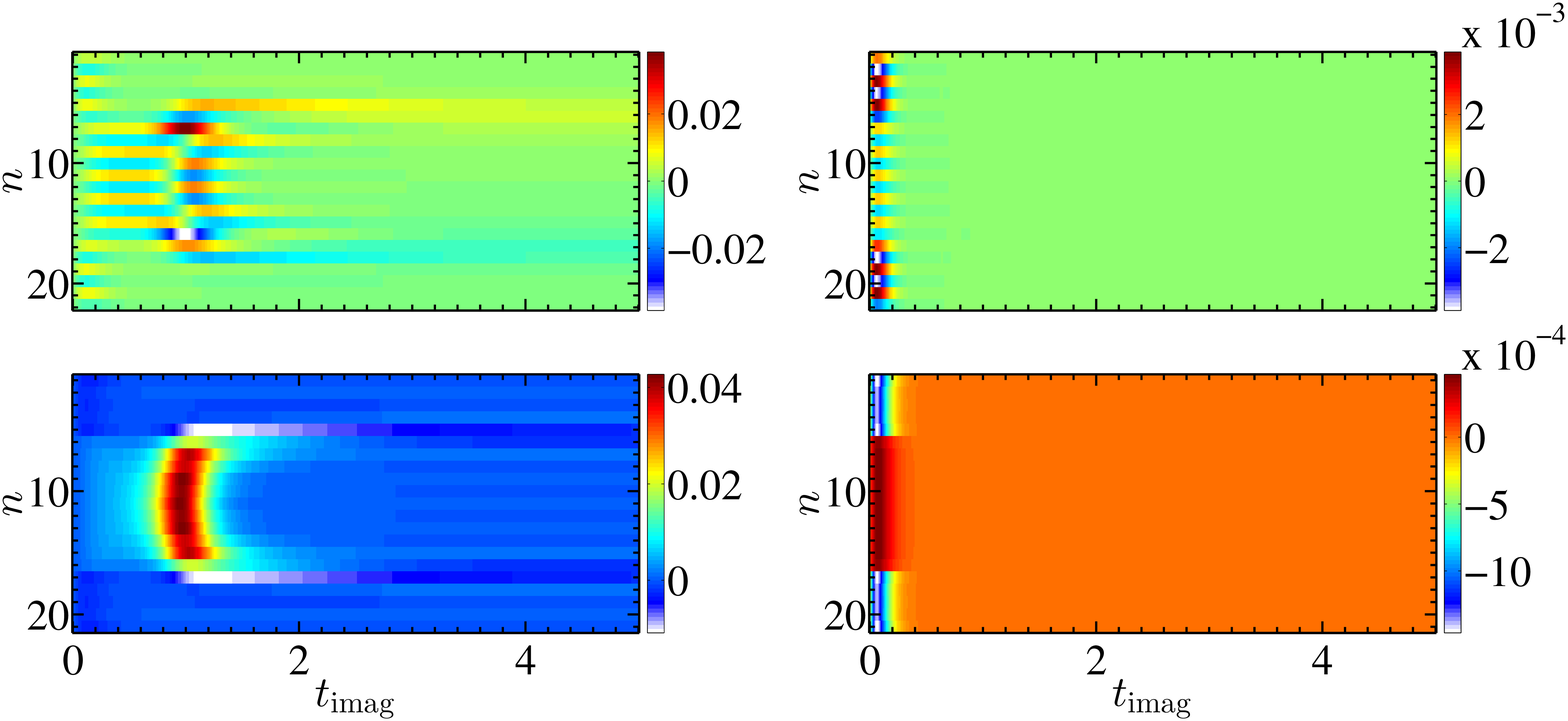}
\includegraphics[width=\columnwidth]{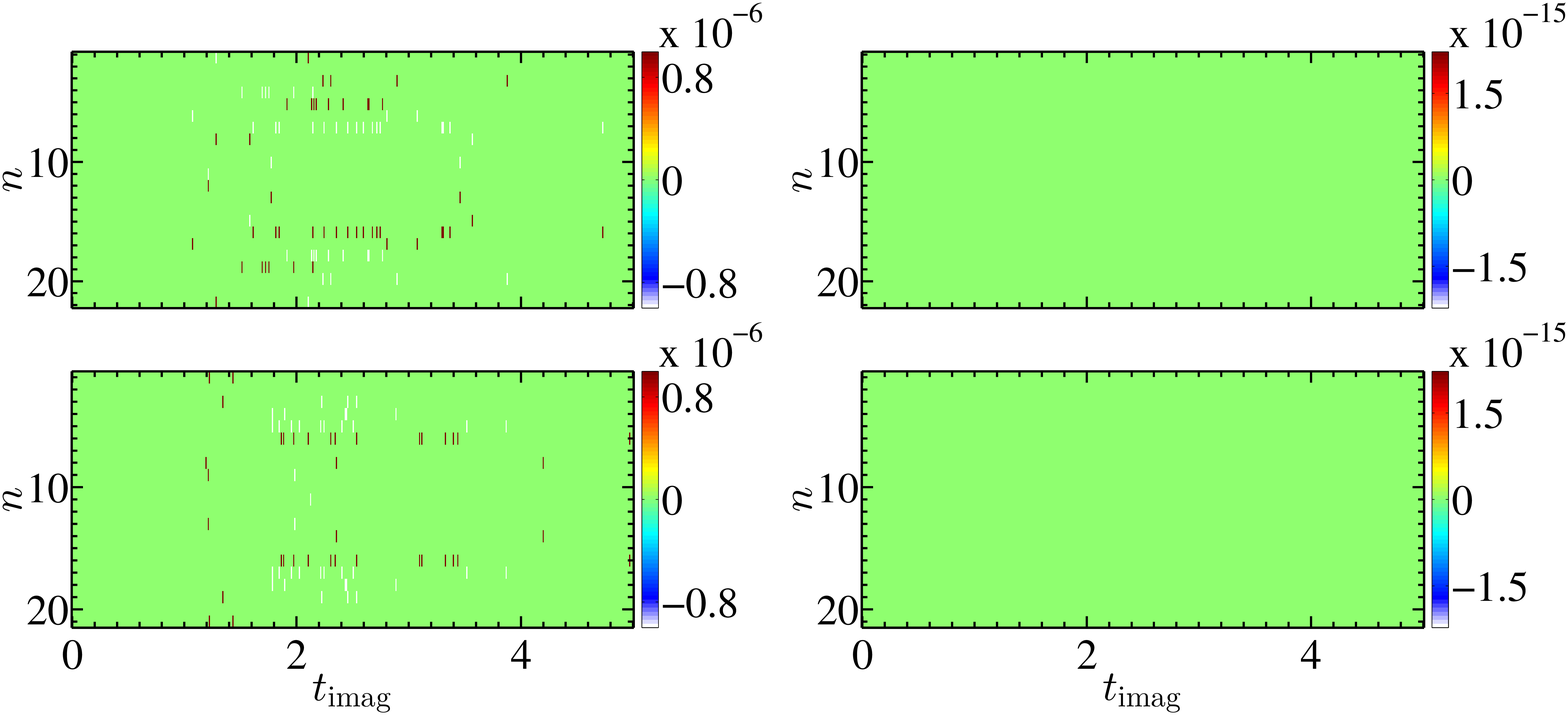}
\caption{Difference in the site resolved expectation values for spin (first and third row) and flux (second and fourth row) for a system of size $N=22$ and dynamical fermion mass $m=3.0$ (left column) and $10.0$ (right column). The upper two rows show the difference between results computed with a time step $\Delta t=\num{1.0d-3}$ and $\num{0.5d-3}$ for $D=100$, the lower two rows the difference between results computed with $D=100$ and $130$ for $\Delta t=\num{1.0d-3}$.}
\label{fig:qq_timag_color_error}
\end{figure}

In figure \ref{fig:qq_timag_color_error} we show how the spin and flux configuration changes for a system of size $N=22$, if a larger bond dimension or a smaller time step is used. As the figure shows, there is no significant improvement with larger bond dimension. Reducing the time step by a factor of two to $\Delta t=\num{0.5d-3}$ leads to a slightly changed spin and flux configuration on the order of $\num{e-2}$ for $m=3.0$. However, these changes are predominantly at an early stage of the evolution and difference in the final configuration is much smaller. For $m=10.0$ the picture is qualitatively similar, at a very early stage there are small differences which are roughly one order of magnitude less than in the $m=3.0$ case. All in all the changes for reduced time step and larger bond dimension are rather small compared to the data for $\Delta t=\num{1.0d-3}$, $D=100$ presented in figure \ref{fig:observables_timag_3D} hence justifying our choice of time step and bond dimension.

\subsection{Real-time evolution with static external charges}
\begin{figure}[!htbp]
\centering
\includegraphics[width=\columnwidth]{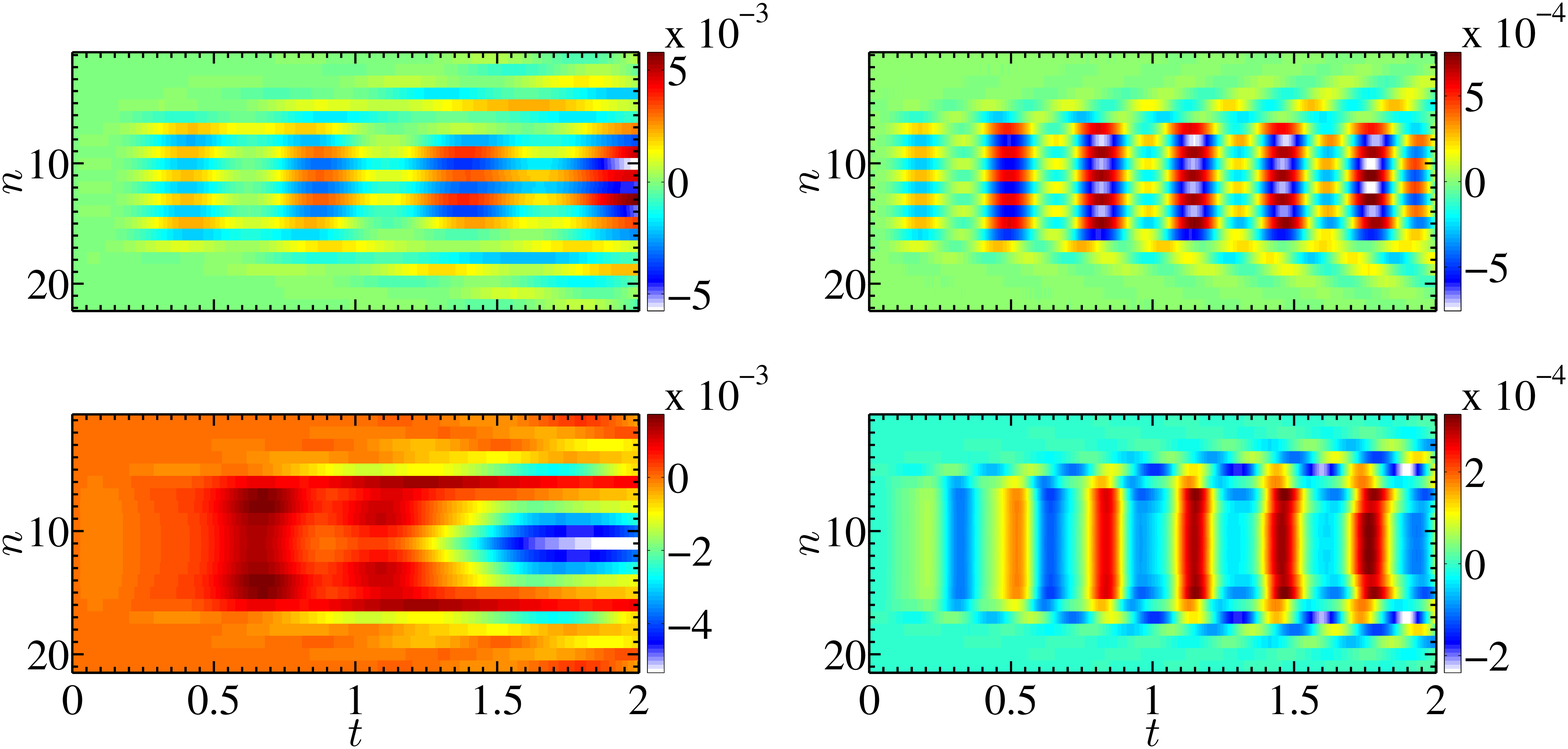}
\includegraphics[width=\columnwidth]{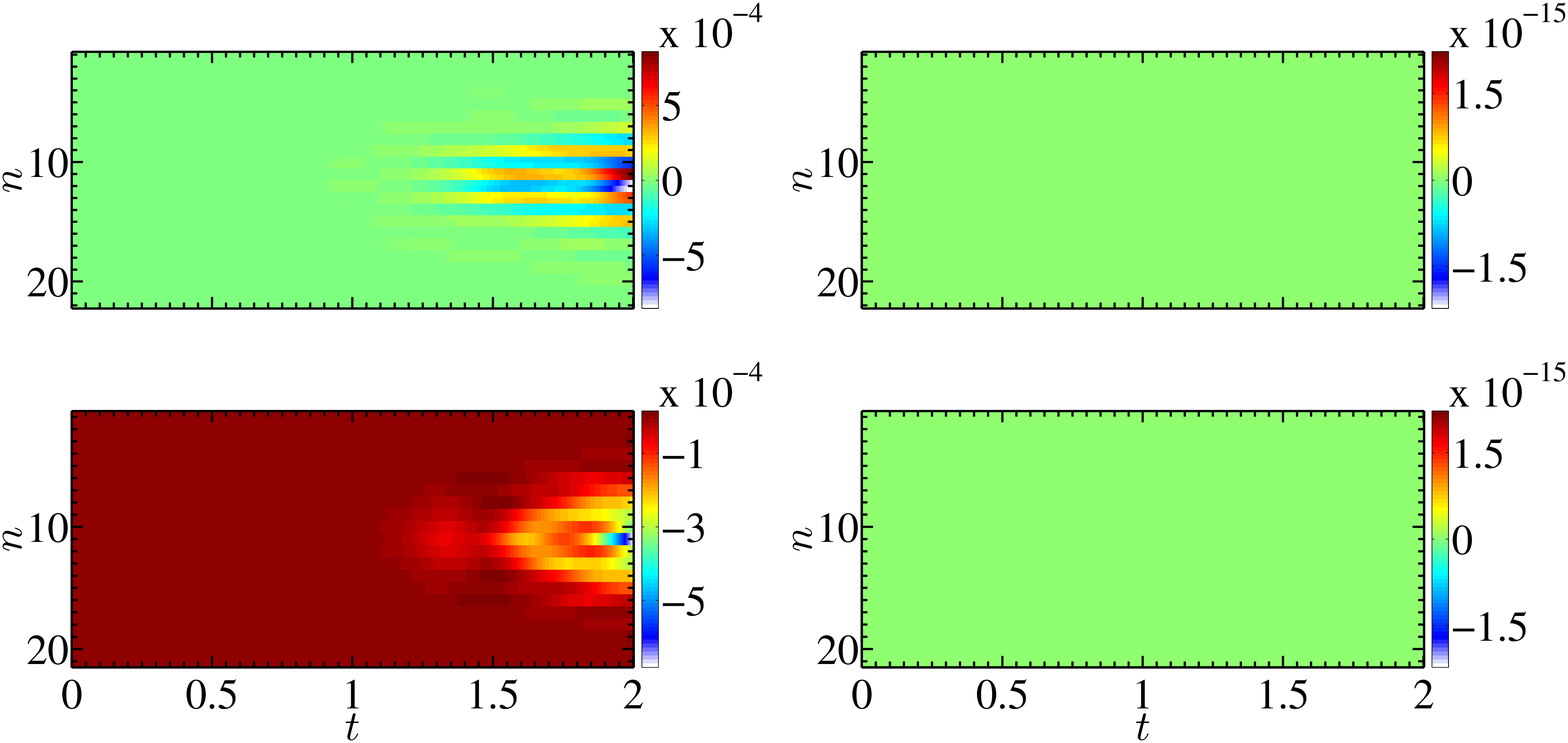}
\caption{Difference in the site resolved expectation values for spin (first and third row) and flux (second and fourth row) for a system of size $N=22$ and dynamical fermion mass $m=3.0$ (left column) and $10.0$ (right column). The upper two rows show the difference between results computed with a time step $\Delta t=\num{1.0d-4}$ and $\num{0.5d-4}$ for $D=100$, the lower two rows the difference between results computed with $D=100$ and $130$ for $\Delta t=\num{1.0d-4}$.}
\label{fig:qq_color_error}
\end{figure}

Also for the real-time evolution case with static external charges, we analyze our error analogous to the imaginary time case. The difference in results for a smaller time step $\Delta t=\num{0.5d-4}$ and larger bond dimension $D=130$ is presented in figure \ref{fig:qq_color_error}. A reduction of the time step shows a rather similar effect to the imaginary time case and leads to differences on the order of $\num{e-3}$ for $m=3.0$ and $\num{e-4}$ for $m=10.0$. For $m=3.0$, the differences comparing to results with enlarged bond dimension are bigger than in the imaginary time case, whereas for $m=10.0$ there is essentially no change up to machine accuracy. Overall, we see the same picture as in the imaginary time cases that our choice of time step and bond dimension controls the error well enough to avoid considerable influences on the effects observed in figure \ref{fig:observables_real_ext_3D}.

\subsection{Real-time evolution with dynamical charges}
\begin{figure}[!htbp]
\centering
\includegraphics[width=\columnwidth]{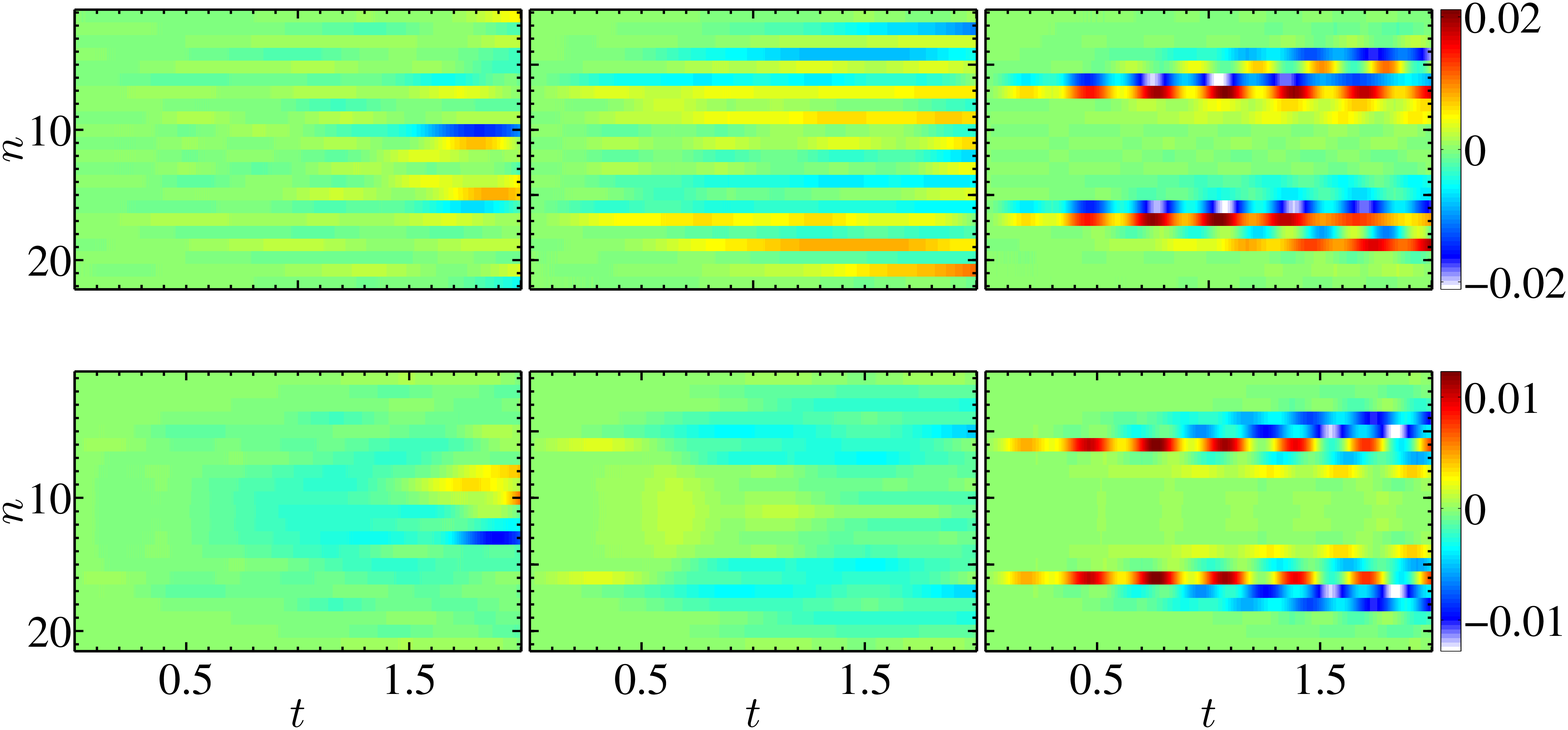}
\includegraphics[width=\columnwidth]{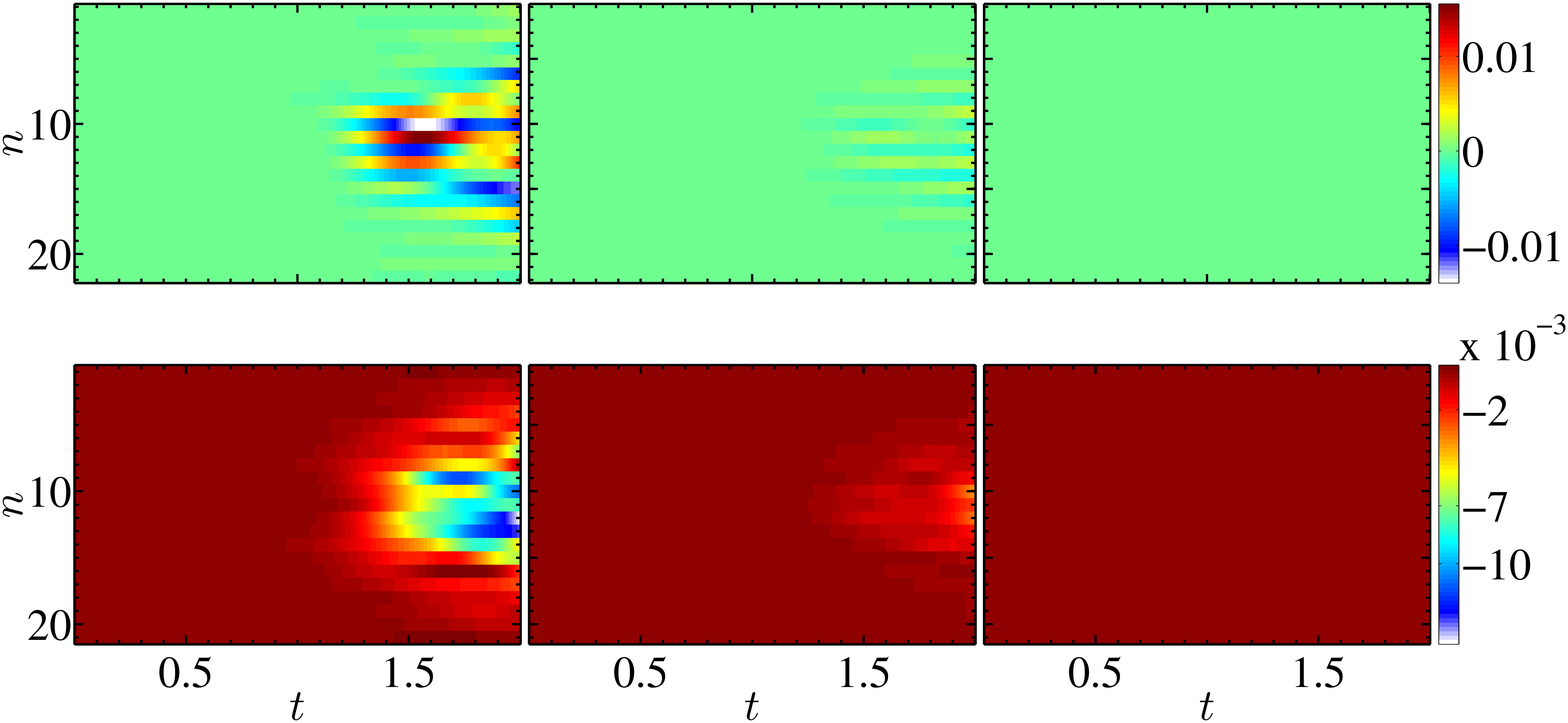}
\caption{Difference of the site resolved expectation values for spin (first and third row) and flux (second and fourth row) for a system of size $N=22$ and dynamical fermion mass $m=1.0$ (left column), $3.0$ (central column) and $10.0$ (right column). The upper two rows show the difference between results computed with a time step $\Delta t=\num{1.0d-4}$ and $\num{0.5d-4}$ for $D=100$, the lower two rows the difference between results computed with $D=100$ and $130$ for $\Delta t=\num{1.0d-4}$.}
\label{fig:string_color_error}
\end{figure}

For the real-time evolution with dynamical charges the same error estimation as in the cases with static charges yields the results in figure \ref{fig:string_color_error}. A reduction of the time step from $\Delta t=\num{1.0d-4}$ to $\num{0.5d-4}$ yields a change in the spin and flux configuration on the order of $\num{e-2}$. Contrary to the cases with static charges, the change is more pronounced for a large fermion mass $m=10.0$, whereas for smaller fermion masses there is less change. For the bond dimension however, we see again the same effect that the changes are more pronounced if the mass is smaller, whereas there is almost no difference for larger mass. Nevertheless, also in this case the absolute changes in the spin and flux configuration are rather small compared to the data presented in figure \ref{fig:observables_real_3D}, thereby showing that the errors due to our choice of time step and bond dimension are negligible.
\end{document}